\documentclass[12pt, draftclsnofoot, onecolumn]{IEEEtran}
\usepackage{amsmath,amsfonts}
\usepackage{amssymb}
\usepackage{array}
\usepackage{changes}
\usepackage[caption=false,textfont=sf]{subfig}
\usepackage{textcomp}
\usepackage{stfloats}
\usepackage{url}
\usepackage{verbatim}
\usepackage{graphicx}
\usepackage{mathtools}
\usepackage{cite}
\usepackage{color}
\usepackage{bm}
\usepackage{xcolor}
\newcommand{\tr}{\mathrm{tr}}
\newcommand{\HH}{\mathrm{H}}	
\newcommand{\TT}{\mathrm{T}}

\newtheorem{theorem}{Theorem}

\newtheorem{remark}{Remark}

\newtheorem{proposition}{Proposition}

\usepackage[ruled,linesnumbered]{algorithm2e}

\begin{document}
	
	\title{Networked Sensing with AI-Empowered Interference Management: Exploiting Macro-Diversity and Array Gain in Perceptive Mobile Networks}
	\author{Lei Xie, Shenghui Song, and Khaled B. Letaief
		\thanks{L. Xie, S. Song, and Khaled B. Letaief  are with Department of Electronic and Computer Engineering, the Hong Kong University of Science and Technology, Hong Kong. e-mail: ($\{$eelxie, eeshsong, eekhaled$\}$@ust.hk).}}
	\maketitle
	\begin{abstract}
		Sensing will be an important service of future wireless networks to assist innovative applications such as autonomous driving and environment monitoring. Perceptive mobile networks (PMNs) were proposed to add sensing capability to current cellular networks. Different from traditional radar, the cellular structure of PMNs offers multiple perspectives to sense the same target, but the inherent interference between sensing and communication along with the joint processing among distributed sensing nodes (SNs) also cause big challenges for the design of PMNs. In this paper, we first propose a two-stage protocol to tackle the interference between two sub-systems. Specifically, the echoes created by communication signals, i.e., interference for sensing, are first estimated in the clutter estimation (CE) stage and then utilized for interference management in the target sensing (TS) stage. 
		A \textit{networked} sensing detector is then derived to exploit the perspectives provided by multiple SNs for sensing the same target. The macro-diversity from multiple SNs together with the array gain and the higher angular resolution from multiple receive antennas of each SN are investigated to reveal the benefit of networked sensing. Furthermore, we derive the sufficient condition to guarantee one SN's contribution is positive, based on which a SN selection algorithm is proposed. To reduce the communication workload, we propose a distributed model-driven deep-learning algorithm that utilizes partially-sampled data for CE. Simulation results confirm the benefits of networked sensing and validate the higher efficiency of the proposed CE algorithm than existing methods.
	\end{abstract}
	
	
	\section{Introduction}
	With the development of innovative applications such as autonomous driving and industrial internet of things (IIoT) \cite{cui2021integrating,10.1145/3447744,9143269}, there is an increasing demand on sensing services such as target positioning, tracking, and environmental monitoring \cite{6504845,9296833}. Unfortunately, the current mobile networks, though very successful in providing communication services, are not able to meet the accurate sensing requirement of future applications. To this end, the recently proposed integrated sensing and communication (ISAC) provides a promising platform to integrate sensing with wireless communication \cite{8999605,9737357}, where the adoption of millimeter wave (mmWave) by 5G and beyond systems further enables the hardware and software reuse between the two systems. As a special type of ISAC, perceptive mobile networks (PMNs) were proposed to integrate sensing ability to current cellular networks \cite{9296833,xie2022collaborative}.
	
	There are many favorable properties of mobile networks that can facilitate sensing. First, the well developed cellular networks can provide a large surveillance coverage. Second, the high-density and multiple-antenna sensing nodes (SNs), such as base stations (BSs), not only offer a sufficient spatial freedom for interference cancellation, but also enables networked sensing. 
	However, there are also new challenges. For example, the integration of sensing and communication requires proper interference management between the two systems in both the device (full-duplex operation \cite{9540344}) and network level (multi-user interference \cite{liu2021integrated}).  
	On the one hand, the newly added sensing signals should avoid generating interference to existing communication users. On the other hand, the interference from communication to sensing, represented by the clutter caused by communication signals\footnote{The echoes backscattered from the clutter patches, including the communication users and their nearby scatter points, are referred to as clutter \cite{1263229,4101326}.}, should also be well handled, for which clutter estimation (CE) is very critical. Finally, networked sensing with distributed nodes may cause a heavy communication and computation workload over the network and faces stringent latency requirement. 
	
	Interference management is at the core of ISAC network design. At the device level, utilizing BSs to serve communication users and sense targets at the same time will cause self-interference and require full-duplex operation. Some research efforts have been made on self-interference cancellation (SIC) \cite{bharadia2013full,sabharwal2014band} to enable full-duplex operation, which unfortunately is still not very mature. The authors of \cite{9296833} addressed the full-duplex issue by separating the sensing transmitter and receiver to different remote radio units (RRUs) in a cloud radio access network (C-RAN). Along the same line of research, the authors of \cite{xie2021perceptive} proposed to utilize another layer of passive target monitoring terminals (TMTs) to save the need for full-duplex operation. 
	
	At the network level, there is an inherent interference between sensing and communication. In particular, in PMNs, communication signals will create the clutter for sensing. 
	The estimation of the clutter is very critical for accurate sensing.
	The authors of \cite{8827589} proposed to construct clutter based on the estimated sensing parameters, e.g., time delay, Doppler frequency and direction, and then remove it from the signal. However, the computational cost of the compressed sensing (CS)-based parameter estimation can be extremely high due to the continuous and rapidly-changing clutter parameters in the space and Doppler domains. This issue will be more serious for networked sensing where information sharing between multiple SNs is necessary. Thus, a computation and communication efficient CE algorithm is desired. 
	
	Although interference management has attracted much attention, networked sensing that can exploit the perspectives from multiple SNs has not been well investigated. In this paper, we will investigate networked sensing and its associated CE. These two issues are similar for PMNs with different SNs, e.g., BSs, RRUs, or TMTs. Here, we consider the PMN with distributed TMTs and the results can be applied to PMNs with other SNs. TMTs are passive nodes with only perception functionalities, including radar, vision, and other sensing capabilities \cite{9143269,7845396}. They are distributed in a target area and connected with the data  center on the base stations (BSs) through low latency links. As a result, BSs will serve as radar transmitters besides performing their communication duty and the sensing task is jointly performed by TMTs to avoid full-duplex operation. 
	
	We will first propose a two-stage protocol where CE and target sensing (TS) are performed in two consecutive periods, respectively, where the clutter created by communication signals will be estimated in the CE period and then utilized for TS. In order to guarantee the clutter statistics do not change in the TS period, the sensing signal is properly designed to avoid affecting the clutter. We then derive a  networked detector based on the generalized likelihood ratio test (GLRT) detection, which is optimal in terms of maximizing the signal-to-clutter-plus-noise ratio (SCNR). Performance analysis shall reveal the impact of several key system parameters, including the number of TMTs and the number of antennas at the TMT. Physical insights with respect to the macro-diversity, array gain, and angular resolution are then revealed, identifying the unique advantages of networked sensing.
	
	To reduce the communication workload for CE, we further propose a distributed clutter covariance estimation algorithm where the estimation is performed at TMTs. The low rank clutter in mmWave channel \cite{sun2013target,7272834,li2016optimum} makes it possible to estimate the clutter covariance by using partial samples of the received signal. However, the estimated covariance matrix may be ill-conditioned due to the limited data samples. To this end, we shall unfold the expectation-maximization (EM) detector with several learnable parameters and propose an EM-Net algorithm, which achieves accurate estimation with less data than existing methods.
	
	The contributions of this paper can be summarized as follows:
	\begin{enumerate}
		\item We propose a two-stage protocol for target sensing in PMNs with the presence of clutter. First, the clutter created by communication signals is estimated in the CE stage, whose results are utilized for interference management in the TS stage. To ensure the clutter in the signal-under-test has the same statistical structure as the estimated one from CE, the precoder for sensing signals is designed in such a way that it does not affect the clutter patches in the TS stage. 
		\item We derive a distributed networked detector where multiple TMTs collaboratively sense a target with a constant false alarm probability. We theoretically evaluate the performance of the proposed detector, whose accuracy is validated by simulation. 
		\item The impact of the number of TMTs and the number of antennas at the TMTs are investigated. For the former, we derive a sufficient condition for the contribution of one TMT to be positive and propose a TMT selection algorithm based on the condition. For the latter, we show that, different from communication,  multiple antennas only provide array gain and higher angular resolution  but no diversity gain, due to the use of only the line-of-sight (LoS) component. 
		\item 
		To improve the communication efficiency, 
		we propose an efficient and distributed CE method by unfolding the EM algorithm where several learnable parameters are introduced. Compared with the existing methods, the proposed algorithm can achieve better estimation performance with less data samples.
	\end{enumerate}
	
	The remainder of this paper is organized as follows. Section II introduces the system model and the two-stage protocol. The networked detector is derived in Section III where its performance evaluation is also given. Based on the analysis results, the impact of several key parameters, including the number of TMTs (macro-diversity) and the number of antennas (array gain and angular resolution) are  investigated. An unfolded EM algorithm is proposed in Section IV for efficient CE. Section V validates the performance of the proposed networked detector and the efficiency of the proposed CE algorithm by simulation. Finally, Section VI concludes this paper.
	
	\section{System Model and Two-stage Sensing}
	
	\begin{figure}[!t]
		\centering
		\includegraphics[width=3.21in]{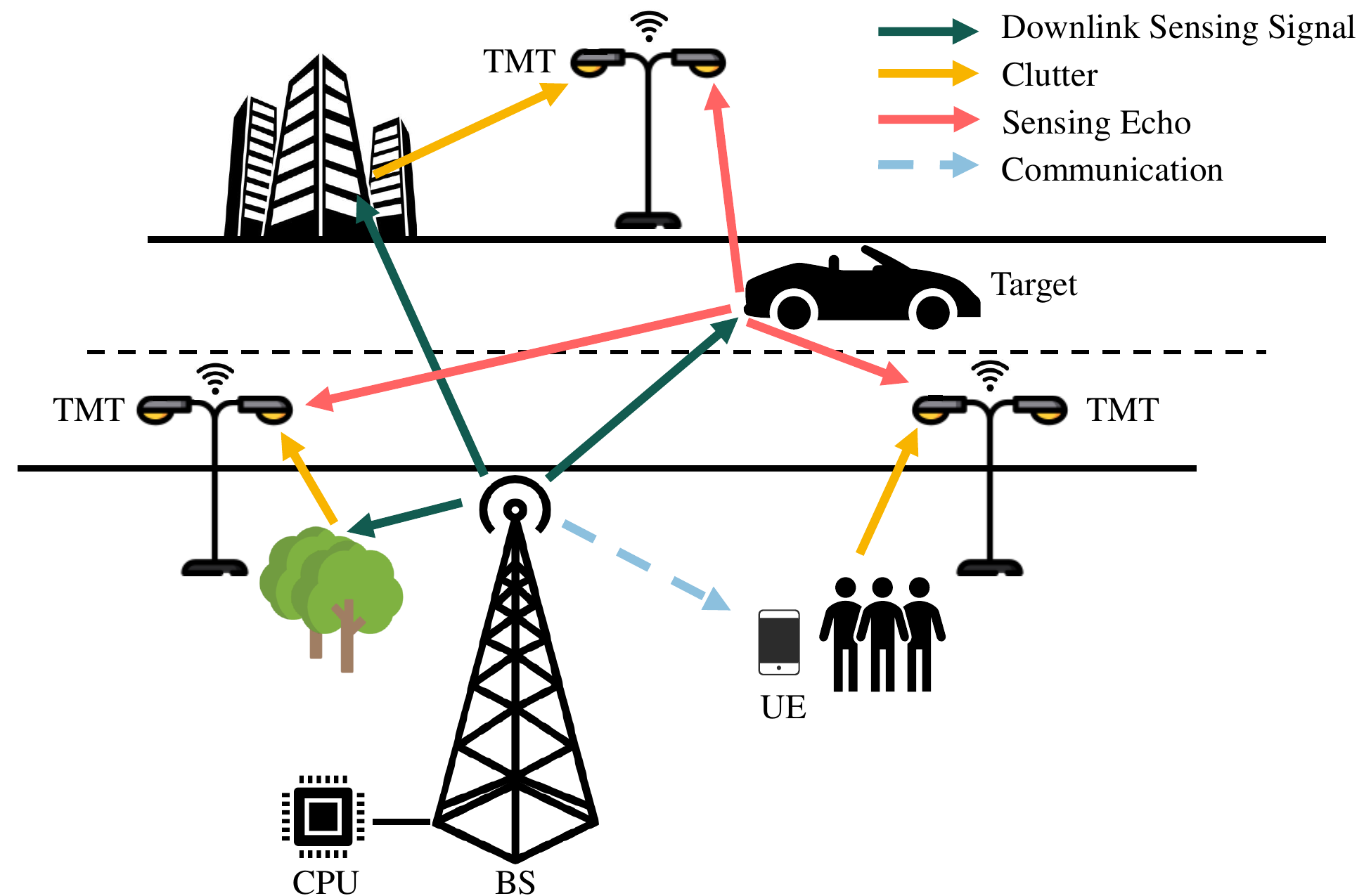}
		\caption{Illustration of the system structure.}
		\label{fig_systemstructure}
	\end{figure}
	
	Consider a PMN where passive TMTs are deployed over conventional cellular networks for sensing purposes, as illustrated in Fig. \ref{fig_systemstructure}. Assume that the BSs and TMTs are equipped with $N_T$ and $N_R$ antennas, respectively. 
	All the TMTs are connected with the data center on the base stations (BSs) through low latency links to achieve clock synchronization, which is an inherent challenge for networked sensing due to the distributed nature of the network \cite{9540344}.
	The objective in this paper is to detect whether a target is present at a given location.
	Note that the addition of the TMTs saves the need for full-duplex operation, while the protocol and algorithms proposed in this paper are valid for other PMN architectures, such as those using full-duplex BSs and RRUs as the SNs.
	The target is assumed to be point-like and static or slow-moving, which is known as the Swerling I model  \cite{9052470}, where the Doppler effect is neglected.
	The target detection problem is formulated as a hypothesis testing between $\mathcal{H}_0$ (target absence) and $\mathcal{H}_1$ (target presence) and  achieved by a likelihood ratio test \cite{1263229,4101326}. The decision statistic requires the statistical information of the clutter to construct the probability density function for both the clutter-alone case ($\mathcal{H}_0$) and the signal-plus-clutter case ($\mathcal{H}_1$).

	To achieve target detection in PMNs, we propose a two-stage protocol as illustrated in Fig. \ref{fig_framestructure}. Note that the normal communication service is not affected by sensing, which only happens in the downlink. In particular, the downlink time is divided into two periods, i.e., the CE period and TS period. BSs only serve the communication users during the CE period, while the radar detection is jointly achieved by multiple TMTs in the TS period. In the following, we explain the detailed operations of the two periods, respectively.  
	
	\subsection{Communication and CE Period}
	In the CE period, the BSs send communication signals to the UEs, which will be reflected by the clutter patches and captured by TMTs for CE. 
	In this paper, we consider targets that are not very close to the UEs. Thus, due to the narrow beam in the mmWave system, the echo reflected from the target can be ignored\footnote{Otherwise, if the echo is not negligible, the effect of the target echo can be alleviated by removing the target signal component from the estimated covariance matrix \cite{8949537} or selecting the target-free data through training sample censoring \cite{6911995}.}.
	Specifically, in the $n$th subframe, the BS transmits communication signals to $K$ UEs and the received signal at the $l$th TMT is given as \cite{6324753}
	\begin{equation}
		\label{yl}
		\begin{split}
			\mathbf{y}_{c,l}(n)=\mathbf{H}_l(n) \mathbf{F} \mathbf{s}(n)+\mathbf{n}_l(n), n=1,2,\cdots,N,
		\end{split}
	\end{equation}
	where $\mathbf{s}(n) \in \mathbb{C}^{K \times 1}$ denotes the communication symbol with covariance matrix $\mathbf{I}$. The precoder matrix $\mathbf{F}\in \mathbb{C}^{N_T \times K}$ can be designed based on some existing methods, e.g., the maximal-ratio combining \cite{win1999analysis} and zero-forcing \cite{wiesel2008zero}, and $\mathbf{n}_l(n)$ is the additive white Gaussian noise (AWGN) with zero mean and covariance matrix $\sigma^2\mathbf{I}$.
	\begin{figure}[!t]
		\centering
		\includegraphics[width=3.21in]{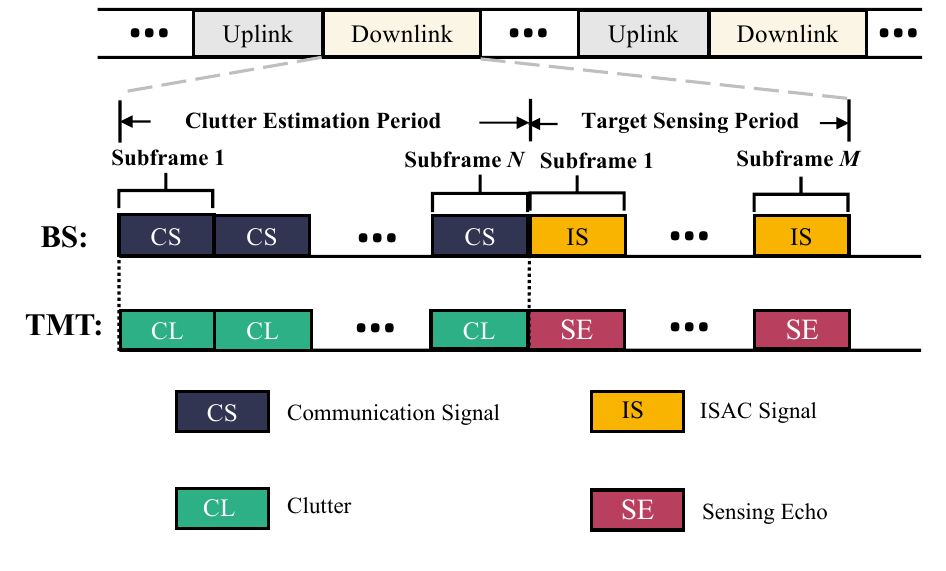}
		\caption{Frame structure for the proposed integrated sensing and communication protocol.}
		\label{fig_framestructure}
	\end{figure}
	Note that $\mathbf{F} \mathbf{s}(n)$ and $\mathbf{y}_{c,l}(n)$ correspond to the ``Communication Signal'' and ``Clutter Echo'' in Fig. \ref{fig_framestructure}, respectively.
	
	Here, $\mathbf{H}_l(n)\in \mathbb{C}^{N_R \times N_T }$ denotes the channel matrix between the BS and the $l$th TMT. With uniform linear array, the steering vector of the BS with angle of departure (AOD) $\phi$ and that of the $l$th TMT with angle of arrive (AOA) $\theta$ are respectively defined as
	\begin{equation}\label{at}
		\mathbf{a}_T(\phi)=\frac{1}{\sqrt{N_T}}\left[1,e^{j2\pi \frac{d}{\lambda}\cos \phi},\cdots,e^{j2\pi (N_T-1)\frac{d}{\lambda} \cos \phi}\right]^\TT\in \mathbb{C}^{N_T \times 1},
	\end{equation}
	\begin{equation}\label{ar}
		\mathbf{a}_{R}(\theta)=	\frac{1}{\sqrt{N_R}}\left[1,e^{j2\pi \frac{d}{\lambda}\cos \theta},\cdots,e^{j2\pi (N_R-1)\frac{d}{\lambda} \cos \theta}\right]^\TT\in \mathbb{C}^{N_R \times 1},
	\end{equation}
	where $d$ is the antenna spacing and $\lambda$ represents the wave length.
	Thus, the channel matrix with $P$ clutter patches is defined as
	\begin{equation}\label{channelmodel}
		\begin{split}
			\mathbf{H}_l(n)=\sqrt{N_R N_T}\sum_{i=1}^{P} \epsilon_{l,i}(n) \mathbf{a}_{R}(\theta_{l,i})\mathbf{a}_T^\HH(\phi_{i})\in \mathbb{C}^{N_R \times N_T},
		\end{split}
	\end{equation}
	where $\epsilon_{l,i}(n)$ denotes the reflecting coefficient of the $i$th clutter patch observed by the $l$th TMT, which is assumed to follow a complex Gaussian distribution with zero mean and variance $\sigma_{l,i}^2$ \cite{akdeniz2014millimeter}.
	$\phi_{i}$ denotes the AOD of the $i$th clutter patch from the BS and $\theta_{l,i}$ represents the
	AOA of the $i$th clutter patch to the $l$th TMT. Here, we omit the paths that are reflected more than one time. 
	These $P$ clutter patches come from two sources: 1) the UEs, and 2) the strong scatter points near UEs. Note that the reflecting coefficient of the UEs may be low, but the transmit power towards the UEs is high. Therefore, the echoes from the UEs can not be neglected. Note that $\mathbf{y}_{c,l}(n), n=1,2,\cdots,N,$ are independent and identically distributed (i.i.d) with respect to $n$, due to the i.i.d. communication signals.

	For ease of illustration, we rewrite (\ref{yl}) as
	\begin{equation}\label{yl2}
		\begin{split}
			\mathbf{y}_{c,l}(n)=\mathbf{A}_{R,l}  \mathbf{t}_l(n)+\mathbf{n}_l(n), n=1,2,\cdots,N,
		\end{split}
	\end{equation}
	where $\mathbf{A}_{R,l} =\left[
	\mathbf{a}_{R}(\theta_{l,1}),\cdots, \mathbf{a}_{R}(\theta_{l,P})
	\right]\in \mathbb{C}^{N_R \times P},$ and \begin{equation}
		\mathbf{t}_l(n) =\left[
	\sqrt{{N_R N_T}}\epsilon_{l,1}(n)\mathbf{a}_T^\HH(\phi_{1})\mathbf{F}\mathbf{s}(n),\cdots, \sqrt{{N_R N_T}}\epsilon_{l,P}(n)\mathbf{a}_T^\HH(\phi_{P})\mathbf{F}\mathbf{s}(n)
	\right]^\TT \in \mathbb{C}^{P \times 1}.
\end{equation}
	Note that the instantaneous value of the reflecting coefficient may change in different frames, but the statistical information of the reflecting coefficient remains constant. Therefore,  $\mathbf{y}_{c,l}(n)$ follows the Gaussian distribution with $\mathbf{y}_{c,l}(n) \sim \mathcal{CN}(\bm{0},\mathbf{R}_{c,l})$ for $n=1,\cdots,N$, where
	\begin{equation}\label{Rcdef}
		\begin{split}
			\mathbf{R}_{c,l}&\triangleq\mathbb{E}\left\{\mathbf{y}_{c,l}(n)\mathbf{y}_{c,l}^\HH(n)\right\}=\underbrace{\mathbf{A}_{R,l} \mathbf{P}_l   \mathbf{A}_{R,l}^\HH}_{\text{Clutter}} + \underbrace{\sigma^2 \mathbf{I}}_{\text{Noise}},
		\end{split}
	\end{equation}
	and $\mathbf{P}_l =\mathrm{Diag}\left(\left[
	\sqrt{{N_R N_T}}\sigma_{l,1}^2|\mathbf{a}_T^\HH(\phi_{1})\mathbf{F}|^2,\cdots, \sqrt{{N_R N_T}}\sigma_{l,P}^2|\mathbf{a}_T^\HH(\phi_{P})\mathbf{F}|^2
	\right]\right)$ with $\mathrm{Diag}(\mathbf{a})$ represents a diagonal matrix whose diagonal is $\mathbf{a}$.
	The main task of the TMTs in this period is to estimate the clutter covariance matrix $\mathbf{R}_{c,l}$ based on the received signal $\mathbf{y}_{c,l}(n), n=1,\cdots,N$, which governs the interference from communication to sensing.

	\subsection{Communication and TS Period}
	In the TS period, the system aims to probe a target without influencing the communication performance. To achieve good sensing performance, the received clutter in the TS period is supposed to have the same second-order statistics as that in the CE period. For that purpose, we need to properly design the precoder of the sensing signal to avoid affecting the covariance structure of the clutter. 
	Assume that the channel information between the BS and UEs, i.e., $\{\phi_{i}\}_{i=1}^P$ are known from channel estimation of the communication systems. 
	To make sure that the second order statistics of the echo signals do not change in the TS period, we need to guarantee that the sensing signal toward the ST, with AoD $\phi_t$, will not create echoes by the clutter patches\footnote{We assume that $\phi_t \neq \phi_{i}, \forall i$, because otherwise the target can not be detected.}.  For that purpose, we define $
	\mathbf{A}_{T} =\left[
	\mathbf{a}_{T}(\phi_{1}),\cdots,\mathbf{a}_{T}(\phi_{P})
	\right]$
	and construct $\mathbf{f}_\perp$ as the projection of $\mathbf{a}_T(\phi_{t})$ in the null-space of $\mathbf{A}_{T}$, i.e., 
	\begin{equation}
		\begin{split}
			\mathbf{f}_\perp=\frac{\mathbf{a}_T(\phi_t)-\mathbf{A}_T(\mathbf{A}_T^\HH \mathbf{A}_T)^{-1}\mathbf{A}_T^\HH \mathbf{a}_T(\phi_t)}{1-\mathbf{a}_T^\HH(\phi_t)\mathbf{A}_T(\mathbf{A}_T^\HH \mathbf{A}_T)^{-1}\mathbf{A}_T^\HH \mathbf{a}_T(\phi_t)}.
		\end{split}
	\end{equation}
	It can be validated that $\mathbf{A}_{T}^\HH\mathbf{f}_\perp=\mathbf{0}$, i.e., $\mathbf{f}_\perp$ will not affect the response on the direction of clutter patches.
	It has been shown in \cite{xie2021perceptive} that redesigning communication signals for sensing is more  efficient than creating a dedicated sensing signal. Thus, we design the precoder and symbols in the TS period as $	\mathbf{F}_{\text{ISAC}}=\mathbf{F}+\mathbf{f}_\perp \bm{\omega}^\TT,$
	where $\bm{\omega}=[\omega_1,\omega_2,\cdots,\omega_K]^\TT$ denotes the weights for the data streams of $K$ UEs\footnote{In general, $\bm{\omega}$ can be determined based on the relation between the direction of target and UEs. Compared with forming a dedicated sensing data stream, the power transmitted to the target by reusing the communication signal is higher \cite{xie2021perceptive}.}. 
	Here, $	\mathbf{F}_{\text{ISAC}} \mathbf{s} = \mathbf{F} \mathbf{s}+\mathbf{f}_\perp \bm{\omega}^\TT \mathbf{s}$
	corresponds to the ``ISAC Signal'' in Fig. \ref{fig_framestructure}. Note that, compared with the transmit signal in the CE period, the additional signal $\mathbf{f}_\perp \bm{\omega}^\TT \mathbf{s}$ will not create echos from the clutter patches, including the UEs. This guarantees that the communication performance will not be affected by sensing and the clutter covariance structure is the same in the CE and TS periods. 
	
	For ease of illustration, we only consider one subframe in the TS period and the result can be extended to the case with more subframes. In this case, the received signal at the $l$th TMT in the TS period can be given by
	\begin{equation}\label{y1l}
		\begin{split}
		\mathbf{y}_{l} &= \left(\sqrt{N_R N_T}\epsilon_{t,l}\mathbf{a}_{R}(\theta_{t,l})\mathbf{a}_T^\HH(\phi_{t})+\mathbf{H}_l\right) 
		\mathbf{F}_{\text{ISAC}} \mathbf{s}+\mathbf{n}_l\\
		&= c_{t,l} \mathbf{a}_{R}(\theta_{t,l})+\mathbf{A}_{R,l}  \mathbf{t}_l+\mathbf{n}_l,
	\end{split}
	\end{equation}
	where $\theta_{t,l}$ denotes the AOA of the target at the $l$th TMT, and
	\begin{equation}\label{ctl}
		\begin{split}
			c_{t,l}=\sqrt{N_R N_T}\epsilon_{t,l}\mathbf{a}_T^\HH(\phi_{t})\mathbf{F}_{\text{ISAC}}\mathbf{s}=\sqrt{N_R N_T}\epsilon_{t,l}\left(\mathbf{a}_T^\HH(\phi_{t})\mathbf{F}\mathbf{s}+\sum_{k=1}^{K} \omega_k s_k\right),
		\end{split}
	\end{equation}
	represents the complex amplitude of the target component with $\epsilon_{t,l}$ denoting the channel coefficient of the BS-target-TMT ($l$th) link. We assume that $\epsilon_{t,l}$ does not change in one TS subframe. 
	Note that $\mathbf{y}_{l}$ corresponds to the “Sensing Echo” in Fig. \ref{fig_framestructure} and the first term in (\ref{y1l}) represents the echo from the target.

	According to (\ref{y1l}), $\mathbf{y}_{l}$ follows a Gaussian distribution with $\mathbf{y}_{l} \sim \mathcal{CN}(c_{t,l}\mathbf{a}_{R}(\theta_{t,l}),\mathbf{R}_{l})$, where
	\begin{equation}\label{muslRsl}
		\begin{split}
			\mathbf{R}_{l}&  \triangleq \mathbb{E}\left\{\left(\mathbf{y}_{l}-c_{t,l} \mathbf{a}_{R}(\theta_{t,l})\right)\left(\mathbf{y}_{l}-c_{t,l} \mathbf{a}_{R}(\theta_{t,l})\right)^\HH\right\}=\mathbf{A}_{R,l} \mathbf{P}_l  \mathbf{A}_{R,l}^\HH + \sigma^2 \mathbf{I}.
		\end{split}
	\end{equation}
	Note that the expectation operations in (\ref{Rcdef}) and (\ref{muslRsl}) are implemented over a whole frame.	By comparing (\ref{Rcdef}) and (\ref{muslRsl}), we can observe that the clutter in the TS period has the same covariance matrix as that in the CE period. Without loss of generality, we denote $\mathbf{R}_{c,l}=\mathbf{R}_{l}$. 
	
	\section{Networked Sensing with Multiple TMTs}
	The distributed TMTs provide multiple perspectives to observe the same target. In this section, we propose a networked detector and then evaluate its performance to reveal some physical insights.
	
	\subsection{Networked Sensing}
	Radar detection is a binary hypothesis testing problem, 
	where hypotheses $\mathcal{H}_0$ and $\mathcal{H}_1$ correspond to the absence and presence of the target, respectively, i.e.,
	\begin{equation}\label{HypPro0}
		\begin{split}
			\mathcal{H}_0:& \mathbf{y}_{l} \sim \mathcal{CN}(\bm{0},\mathbf{R}_l)\\
			\mathcal{H}_1:& \mathbf{y}_{l} \sim \mathcal{CN}(c_{t,l}\mathbf{a}_{R}(\theta_{t,l}),\mathbf{R}_l)
		\end{split}
	\end{equation}
	with $\mathbf{y}_{l}$ denoting the signal-under-test at the $l$th TMT in the TS period. 
	The conditional probability density function (pdf) of $\mathbf{y}_{l}$ under two hypotheses are given by
	\begin{equation}\label{ylpdf}
		\begin{split}
			&f (\mathbf{y}_{l}|\mathcal{H}_0)=C_{N} \det(\mathbf{R}_{l})^{-1}\exp(-\mathbf{y}_{l}^\HH \mathbf{R}_{l}^{-1}\mathbf{y}_{l}),\\
			&f (\mathbf{y}_{l};c_{t,l}|\mathcal{H}_1)=C_{N} \det(\mathbf{R}_l)^{-1}\exp(-(\mathbf{y}_{l}-c_{t,l}\mathbf{a}_{R}(\theta_{t,l}))^\HH \mathbf{R}_l^{-1}(\mathbf{y}_{l}-c_{t,l}\mathbf{a}_{R}(\theta_{t,l}))),
		\end{split}
	\end{equation}
	where $C_{N}$ is a normalization coefficient. 
	Given the echo signals received by different TMTs are independent due to the independent reflecting coefficients, 
	the optimal detector that maximizes the output signal-to-clutter-plus-noise ratio (SCNR)
	is the generalized likelihood ratio test (GLRT) detector \cite{135446}, i.e., 
	\begin{equation}\label{AMFdef}
		\begin{split}
			\Delta_L=\frac{\max\limits_{\{c_{t,l}\}} \prod\limits_{l=1}^L f (\mathbf{y}_{l};c_{t,l}|\mathcal{H}_1) }{\prod\limits_{l=1}^Lf (\mathbf{y}_{l}|\mathcal{H}_0)}  \mathop{\gtrless}\limits_{\mathrm{H}_0}^{\mathrm{H}_1} \delta_L,
		\end{split}
	\end{equation}
	where $\delta_L$ denotes the detection threshold. 	By taking the logarithm on $\Delta_L$, we have the log-likihood ratio
	\begin{equation}\label{AMFlogtest}
		\begin{split}
			\log\Delta_L =\sum_{l=1}^{L}  2\Re(c_{t,l}^* \mathbf{a}_{R,l}^\HH(\theta_{t,l}) \mathbf{R}_l^{-1}\mathbf{y}_{l})-|c_{t,l}|^2 \mathbf{a}_{R,l}^\HH(\theta_{t,l}) \mathbf{R}_l^{-1}\mathbf{a}_{R,l}(\theta_{t,l}),
		\end{split}
	\end{equation}
	where $\Re(\cdot)$ takes the real part of a complex number.
	Maximizing $\log\Delta_L$ with respect to the unknown complex amplitude $c_{t,l}$ by setting $\frac{\partial \log\Delta_L}{\partial c_{t,l}}=0$ yields
	\begin{equation}\label{ctlest}
		\begin{split}
			\hat{c}_{t,l}=\frac{\mathbf{a}_{R,l}^\HH(\theta_{t,l}) \mathbf{R}_l^{-1}\mathbf{y}_{l}}{\mathbf{a}_{R,l}^\HH(\theta_{t,l}) \mathbf{R}_l^{-1}\mathbf{a}_{R,l}(\theta_{t,l})}.
		\end{split}
	\end{equation}
	By substituting (\ref{ctlest}) into (\ref{AMFlogtest}), the decision statistic of the joint GLRT detector is given as
	\begin{equation}
		\begin{split}\label{AMFresult}
			\Gamma=\sum_{l=1}^{L} \Gamma_l\mathop{\gtrless}\limits_{\mathrm{H}_0}^{\mathrm{H}_1} \gamma_L,
		\end{split}
	\end{equation}
	where 
	\begin{equation}\label{gammal}
		\begin{split}
			\Gamma_l\triangleq \frac{|\mathbf{a}_{R}^\HH(\theta_{t,l}) \mathbf{R}_l^{-1}\mathbf{y}_{l}|^2}{\mathbf{a}_{R}^\HH(\theta_{t,l}) \mathbf{R}_l^{-1}\mathbf{a}_{R}(\theta_{t,l})}
		\end{split}
	\end{equation}
	and	$\gamma_L =\log \delta_L$ denotes the detection threshold for $\Gamma$.
	
	For ease of illustration, we will utilize $\mathbf{a}_{t,l}$ to denote $\mathbf{a}_{R}(\theta_{t,l})$.	
	Under $\mathcal{H}_0$, the signal-under-test only contains clutter and noise. 
	Thus, $\Gamma_l$ in (\ref{gammal}) follows a central chi-square distribution with 2 degrees of freedom (DOF), i.e., $\Gamma_l\sim \chi_{2}^2 \left( 0 \right)$.
	Under $\mathcal{H}_1$, $\Gamma_l$ follows a chi-square distribution with 2 degrees of freedom (DOF) and non-central parameter $\mu_l^2$, i.e., $\Gamma_l \sim \chi_{2}^2 \left( \mu_l^2 \right)$,
	where $\mu_l^2=c_{t,l}^2 \mathbf{a}_{t,l}^\HH\mathbf{R}_l^{-1}\mathbf{a}_{t,l}$.
	Thus, we have
	\begin{equation}\label{GammaDet0}
		\begin{split}
			\Gamma=\sum_{l=1}^L \Gamma_l \sim \left\{ 
			\begin{matrix}
				\chi_{2L}^2 \left( \zeta_L \right),& \mathcal{H}_1,\\
				\chi_{2L}^2 \left( 0 \right), &\mathcal{H}_0,\\
			\end{matrix}
			\right.
		\end{split}
	\end{equation}
	where $\zeta_L=\sum_{l=1}^{L}\mu_l^2$ is the non-central parameter of the chi-square decision statistic $\Gamma$. 
	
	Following the result in  \cite{kelly1989adaptive}, the false alarm probability is given by	 $	P_{fa}=e^{-\frac{\gamma_L}{2}} \sum_{l=0}^{L-1}\frac{\left(\frac{\gamma_L}{2}\right)^l}{l!},$
	indicating that $P_{fa}$ depends on $\gamma_L$ and $L$, but is independent of the clutter covariance $\{\mathbf{R}_l\}$. The decision threshold can be determined without prior knowledge of the clutter, i.e., \cite{32133}
	\begin{equation}\label{threshold}
		\begin{split}
			\gamma_L \approx \left(L-\frac{1}{2}\right)+\left(\sqrt{-\frac{8}{5}\ln (4P_{fa}(1-P_{fa}))} +\sqrt{L-\frac{1}{2}}\right)^2,
		\end{split}
	\end{equation}
	which is also known as the constant false alarm rate (CFAR) property \cite{135446,9052470}. Meanwhile, based on (\ref{threshold}), the false alarm probability $P_{fa}$ remains a constant as $L$ increases. These properties improve the robustness of the system. 
	Finally, the detection probability can be written as \cite{omura1965some}
	\begin{equation}\label{AMFl1}
		\begin{split}
			P_d^{(L)}=Q_{L}\left(\sqrt{\zeta_L},\sqrt{\gamma_L }\right)=Q_{L}\left(\sqrt{\sum_{l=1}^{L}c_{t,l}^2 \mathbf{a}_{t,l}^\HH\mathbf{R}_l^{-1}\mathbf{a}_{t,l}},\sqrt{\gamma_L }\right),
		\end{split}
	\end{equation}
	where $Q_{k}\left(\cdot,\cdot\right)$ denotes the generalized Marcum Q function of order k.
	In the following, we analyze the impact of the two most important system parameters, namely, the number of antennas at each TMT and the number of TMTs participating in the networked sensing.
	
	\subsection{Array Gain and Angular Resolution: Impact of multiple antennas in one TMT}
	In this section, we will investigate the contribution of one multi-antenna TMT through analyzing the non-central parameter $\mu_l^2$. For that purpose, we first perform the eigen-decomposition on $\mathbf{R}_{l}$ with $\mathbf{R}_l=\mathbf{V}_{l} \mathbf{\Lambda}_{l} \mathbf{V}_{l}^\HH+\sigma^2 \mathbf{I},$
	where $	\mathbf{V}_{l}=\left[\mathbf{v}_{l,1},\cdots,\mathbf{v}_{l,r_l}\right]\in \mathbb{C}^{N_R\times r_l}$ and $	\mathbf{\Lambda}_{l}=\mathrm{Diag}\left(\lambda_{l,1},\cdots,\lambda_{l,r_l}\right)$ with $r_l\triangleq \mathrm{rank}(\mathbf{A}_{R,l}\mathbf{P}_l\mathbf{A}_{R,l}^\HH)$.
	Therefore, we have $	\mathbf{R}_l^{-1}
	=\frac{1}{\sigma^2}\left( \mathbf{I}- \mathbf{V}_{l}( {\sigma^2_n} \mathbf{\Lambda}_{l}^{-1}+  \mathbf{V}_{l}^\HH \mathbf{V}_{l}) ^{-1}  \mathbf{V}_{l}^\HH\right),$
	where we have used the matrix inversion lemma. In the high clutter to noise ratio (CNR) regime \cite{1263229,4101326}, i.e., $\lambda_{l,i} \gg\sigma^2$, we have
	\begin{equation}\label{vsproj}
		\mu_l^2=c_{t,l}^2  \mathbf{a}_{t,l}^\HH\mathbf{R}_l^{-1}\mathbf{a}_{t,l}\approx \frac{c_{t,l}^2}{\sigma^{2}}  ||\mathbf{P}_{V,l}^\perp \mathbf{a}_{t,l}||^2=\text{SNR}_l \cdot \cos^2 \vartheta_{tv,l}^\perp,
	\end{equation}
	where $\text{SNR}_l\triangleq \frac{c_{t,l}^2}{\sigma^{2}}$ denotes the signal-to-noise ratio (SNR) at the $l$th TMT. Here, 	$\mathbf{P}_{V,l}^{\perp}=\mathbf{I}-\mathbf{V}_{l}(   \mathbf{V}_{l}^\HH \mathbf{V}_{l}) ^{-1}  \mathbf{V}_{l}^\HH$
	denotes the projector onto the null space of $\text{span}(\mathbf{V}_{l})$, and $\vartheta_{tv,l}^\perp$ represents the angle between $\mathbf{P}_{V,l}^{\perp}\mathbf{a}_{t,l}$ and $\mathbf{a}_{t,l}$. 
	We can rewrite $\text{SNR}_l$ as $\text{SNR}_l=\frac{C_g N_R P_T}{r_l^\beta}$, 
	where $C_g$ is a constant related to the noise variance, reflection coefficient, and antenna gains, $P_T$ represents the transmission power, $r_l$ denotes the length of the BS-target-TMT link for the $l$-th TMT, and $\beta$ is the path loss exponent.
	Meanwhile, we have 
	$\cos^2 (\vartheta_{tv,l}^\perp)= 1-\cos^2 (\vartheta_{tv,l})=1-|\mathbf{a}_{t,l}^\HH \mathbf{a}_{p,l}|,$
	where $\mathbf{a}_{p,l}\triangleq \mathbf{P}_{V,l} \mathbf{a}_{t,l}$ denotes the orthogonal projection of $\mathbf{a}_{t,l}$ onto $\text{span}(\mathbf{V}_{l})$. According to (\ref{Rcdef}), we have $\text{span}\left(\mathbf{V}_{l}\right)=\text{span}\left(\mathbf{A}_{R,l}\right)$.
	As a result, there exists a set of positive weights $\{\alpha_{i,l}\in [0,1]\}$, such that $\mathbf{a}_{p,l}=\sum_{i=1}^{P} \alpha_{i,l} \mathbf{a}_R(\theta_{l,i})$. It thus follows from (\ref{ar}) that 
	\begin{equation}\label{atpl1}
		\begin{split}
			|\mathbf{a}_{t,l}^\HH \mathbf{a}_{p,l}|&=\frac{1}{N_R}\left\vert 
			\sum_{i=1}^{P}\sum_{n=1}^{N_R}\alpha_{i,l} e^{j2\pi(n-1)\frac{d}{\lambda}\left(\cos \theta_{t,l}-\cos  \theta_{l,i} \right)}\right\vert\\
			&=\frac{1}{N_R}\left\vert \sum_{i=1}^{P} \alpha_{i,l}\frac{e^{j\pi N_R \frac{d}{\lambda}\left(\cos \theta_{t,l}-\cos  \theta_{l,i} \right)}\sin\left(\pi N_R \frac{d}{\lambda}\left(\cos \theta_{t,l}-\cos  \theta_{l,i} \right)\right)}{e^{j\pi \frac{d}{\lambda}\left(\cos \theta_{t,l}-\cos  \theta_{l,i} \right)}\sin\left(\pi  \frac{d}{\lambda}\left(\cos \theta_{t,l}-\cos  \theta_{l,i} \right)\right)}\right\vert,
		\end{split}
	\end{equation} 
	where we have utilized the Euler’s identity, i.e., $2j \sin \theta = e^{j\theta}-e^{-j\theta}$.
	
	Substituting 
	(\ref{atpl1}) into (\ref{vsproj}) yields
	\begin{equation}\label{varsigma00}
		\begin{split}
			\mu_l^2 &\approx 
			\frac{C_g N_R P_T}{r_l^\beta}
			\left(1-
			\left\vert \sum_{i=1}^{P}  \alpha_{i,l}e^{j\pi (N_R-1) \frac{d}{\lambda}\left(\cos \theta_{t,l}-\cos  \theta_{l,i} \right)} \frac{\text{sinc}\left( N_R \Delta_{l,i}\right)}{\text{sinc}\left( \Delta_{l,i}\right)}\right\vert\right),
		\end{split}
	\end{equation} 
	where $\text{sinc}(x)=\frac{\sin \pi x}{\pi x}$ and $	\Delta_{l,i}=\frac{d}{\lambda}\left(\cos \theta_{t,l}-\cos \theta_{l,i}\right). $
	
	\begin{remark}\label{re1}
		It can be observed from (\ref{varsigma00}) that the contribution of the $l$th TMT is determined by several parameters. 
		\begin{enumerate}
			\item The length of the BS-Target-TMT link, $r_l$, affects the SNR exponentially.
			\item The relation between $\theta_{t,l}$ and $\{\theta_{l,i}\}_{i=1}^P$ affects the ability of the $l$th TMT to suppress the clutter. 
			To obtain a larger $\mu_l^2$, we want the summation in (\ref{varsigma00}) to be small.	As a result, $\theta_{t,l}$ and $\theta_{l,i}$ are preferred to be far apart, i.e., a TMT with ``clearer'' view of the target is preferred.  
			\item The number of antennas at the TMT has two effects.
			On the one hand, $\text{SNR}_l$ is directly proportional to $N_R$, which comes from the antenna \emph{array gain}. On the other hand, $\cos \vartheta_{tv,l}$ also depends on $N_R$, which is referred to as the \emph{angular resolution} of the TMT. In particular, the mainlobe of $\text{sinc}\left( N_R \Delta\right)$ can be obtained by setting $\pi N_R  \Delta=\pi$, which gives the boundary of the mainlobe at $\Delta_{ml} = \frac{1}{N_R}$. For the considered ULA, $\text{sinc}\left( N_R \Delta\right)$ is approximately $13$ dB down from the peak of the mainlobe when $\Delta$ is out of $(-\Delta_{ml}/2,\Delta_{ml}/2)$.
			For given $\theta_{t,l}$ and $\theta_{l,i}$, $\Delta_{ml}$ will decrease as $N_R$ increases and the mainlobe of $\text{sinc}\left( N_R \Delta\right)$ will become narrower, leading to a larger $\mu_l^2$. 
		\end{enumerate}
	\end{remark}
	
	\begin{remark}
		The impact of multiple antennas in sensing is different from that in communication.  For instance, multiple antennas can offer diversity gain in wireless communications. However, for sensing, only the LoS component is utilized and the non-line-of-sight (NLoS) components are regarded as part of the clutter. As a result, no diversity gain is provided by multiple receive antennas in sensing applications and $\Gamma_l$ in (\ref{gammal}) only has one complex DOF. However, a larger number of antennas does provide a higher array gain and higher angular resolution, which lead to larger $\text{SNR}_l$ and $\cos^2 \vartheta_{tv,l}^\perp$, respectively.  
	\end{remark}
	
	\subsection{Macro-diversity: Contribution of multiple TMTs}
	In this section, we investigate the benefit of networked sensing. 
	
	\subsubsection{Impact of the number of TMTs}
	It follows from (\ref{AMFl1}) that the detection probability $P_d^{(L)}$ depends on $L$, $\zeta_L$, and $\gamma_L$. The collaboration of multiple TMTs will provide more perspectives for a given target. For instance, when $L$ is larger, it is more likely to find a pair of $\theta_{t,l}$ and $\theta_{l,i}$ which are far apart. However, the detection probability is not a monotonic increasing function of L. Assume there are already $L$ activated TMTs with detection probability $P_d^{(L)}$. Let $P_d^{(L+1)}$ denote the detection probability when a new TMT is selected. In the following, we give a sufficient condition for the contribution of the $(L+1)$th TMT to be positive.
	
	\begin{proposition}\label{Prop1}
		For a fixed false alarm probability $P_{fa}$, we have $P_d^{(L+1)}>P_d^{(L)}$, if the following conditions are satisfied:
		\begin{enumerate}
			\item The non-central parameter with $L$ TMTs is greater than the detection threshold, i.e., $\zeta_{L}>\gamma_{L}$;	
			\item The contribution of the $(L+1)$th TMT is greater than the increment of the threshold, i.e.,
			\begin{equation}\label{cond2}
				\mu_{L+1}^2 \geq \gamma_{L+1}-\gamma_{L}=2\sqrt{-\frac{8}{5}\ln (4P_{fa}(1-P_{fa}))}\cdot\left(\sqrt{L+\frac{1}{2}}-\sqrt{L-\frac{1}{2}}\right)+2.
			\end{equation}
		\end{enumerate}
	\end{proposition}
	
	\emph{Proof:} See Appendix \ref{Prop1proof}.
	
	\begin{remark} \label{remark2}
		In networked sensing, adding one more TMT will change the distribution of the decision statistics under both hypotheses $\mathcal{H}_0$ and $\mathcal{H}_1$. Thus, for a given false alarm probability, one more TMT will lead to a higher detection threshold, and not necessarily provide a higher detection probability. Consider an extreme case when the link between the target and the $(L+1)$th TMT is totally blocked. Under such circumstances, what the new TMT can contribute is only noise, causing a worse probability of detection. \emph{Proposition \ref{Prop1}} provides the condition with which the contribution of the $(L+1)$th TMT is positive. 
	\end{remark}
	
	\subsubsection{TMT Selection Algorithm}
	In practice, many TMTs may be around and it is unnecessary and even harmful to activate all TMTs to sense one target. Thus, the selection of TMTs is critical for networked sensing. One application of \emph{Proposition \ref{Prop1}} is for TMT selection. Assume there are $Q$ TMTs available in an effective area around the target. We propose a selection algorithm, as summarized in Algorithm \ref{alg_selection}.  In particular, we first calculate $\mu_l^2$ for all available TMTs and order them in descending order. Then, the TMTs are selected based on the condition in \emph{Proposition \ref{mono00}}, i.e., we keep adding new TMTs until the condition no longer holds. 
	
	\begin{algorithm}[t] 
		\caption{TMT selection algorithm} 
		\begin{enumerate}
			\item Initialize $n=0$, $\mathcal S=\emptyset$, and $\mathcal U=\{1,2,\cdots,M\}$. 
			\item Find the $i_n$th TMT from $\mathcal U$ that maximizes $\mu_{i_n}^2$, and update $\mathcal S=\mathcal S\cup \{ i_{n} \}$, $\mathcal U= \mathcal U\cap \mathcal S^c$.
			\item $n\gets n+1$.
			\item Repeat 2) to 3) until the conditions in \emph{Proposition \ref{mono00}} is not met.
		\end{enumerate}
		\label{alg_selection} 
	\end{algorithm}

	\section{AI-Empowered Clutter Estimation}
\begin{figure}[!t]
		\centering
		\includegraphics[width=3.21in]{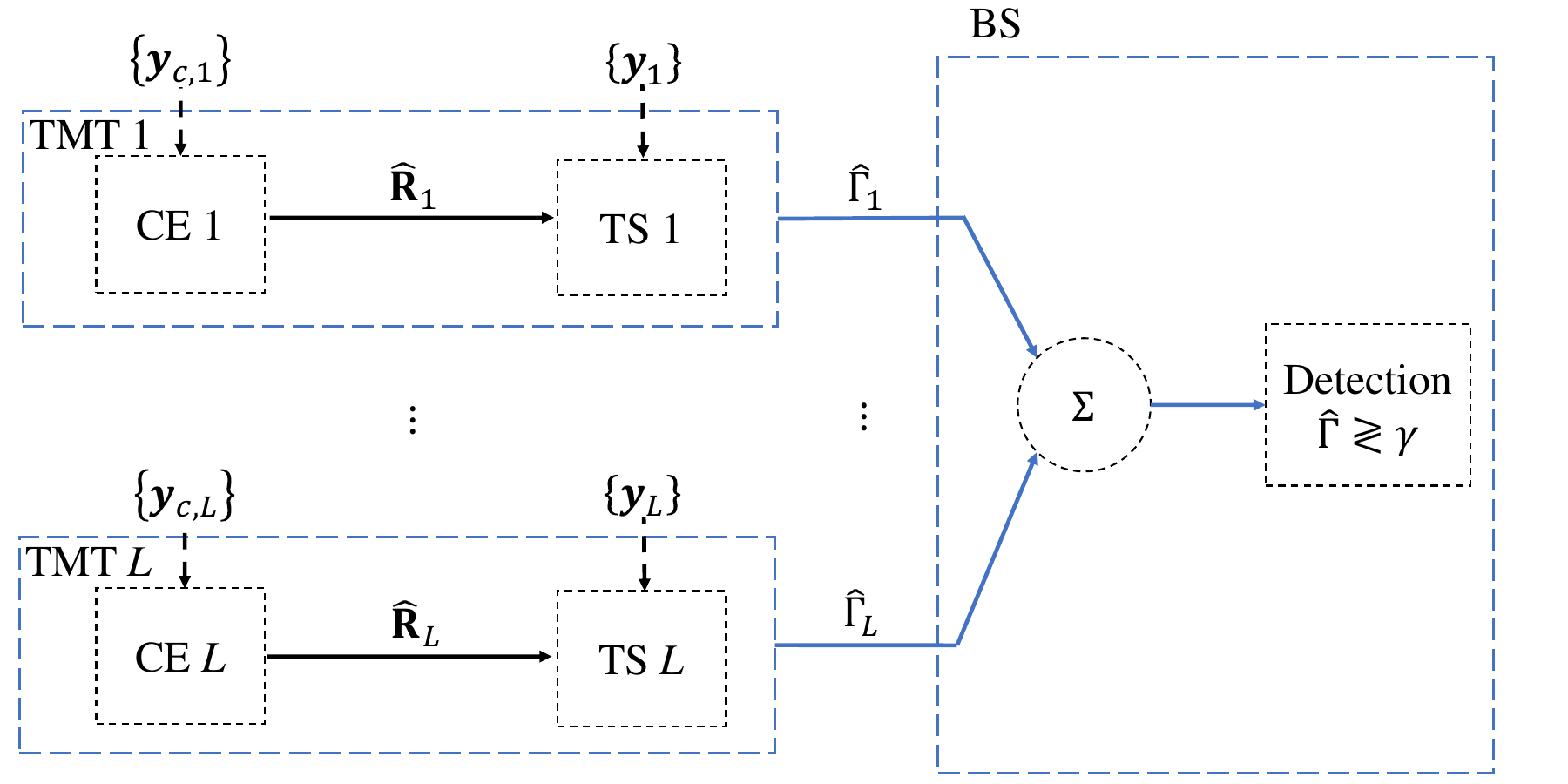}
		\caption{The diagram of networked sensing.}
		\label{fig_sys_diagram}
	\end{figure}

	The networked detector needs to know the second order statistics of the clutter, i.e., the covariance matrices $\{\mathbf{R}_l\}_{l=1}^L$. Unfortunately, they are unknown in real applications and are normally replaced by their estimates $\{\widehat{\mathbf{R}}_l\}_{l=1}^L$. Fig. \ref{fig_sys_diagram} illustrates the diagram of networked sensing, where the covariance matrices $\{\widehat{\mathbf{R}}_l\}_{l=1}^L$ are estimated based on the training samples in the CE period, i.e., $\{\mathbf{y}_{c,l}\}$, and utilized for target sensing, based on the received signals of the TMTs in the TS period, i.e., $\{\mathbf{y}_{l}\}$. 
	
	For the networked sensing considered in this paper, there are issues for CE from both communication and computation perspectives.
	Estimating $\{\widehat{\mathbf{R}}_l\}_{l=1}^L$ for all TMTs by the BS is challenging because moving data from TMTs to the BS can cause very heavy communication burden and lead to serious latency. One possible solution is to estimate $\{\widehat{\mathbf{R}}_l\}_{l=1}^L$ at the TMTs to avoid the heavy communication workload. 
	Unfortunately, the widely-used estimation method, i.e., the sample covariance matrix (SCM) based approach \cite{8999605}, requires a large number samples to guarantee a considerable performance, which cost high hardware resource and power consumption.
	However, the TMTs in the PMN are normally  power-limited. 
	It is thus necessary to develop some sample-efficient algorithms.
	Fortunately, the low-rank of the clutter in the mmWave band makes it possible to estimate the covariance matrix by a small amount of data, which can significantly reduce the workload of TMTs. In this section, we propose an efficient and distributed covariance estimation algorithm based on partial data. 
	
	\subsection{Clutter Estimation with Partial Data}
	To reduce the computation workload, the TMTs sample the clutter echo to obtain
	\begin{equation}\label{HypProOut}
		\begin{split}
			\mathbf{p}_{c,l}(n) =\mathbf{\Omega}_{l,n} \mathbf{y}_{c,l}(n) \in \mathbb{C}^{p_n \times 1}, n=1,\cdots,N,
		\end{split}
	\end{equation}
	where $\mathbf{\Omega}_{l,n} \in \mathbb{C}^{p_n \times N_R}$ denotes the sampling matrix  at the $l$th TMT in the $n$th subframe. In particular, if the $(i,j)$th entry of $\mathbf{\Omega}_{l,n}$ equals to 1, then the $j$th entry of $\mathbf{y}_{c,l}(n)$ is selected as the $i$th entry of $\mathbf{p}_{c,l}(n)$. There is only one ``1'' in each row of $\mathbf{\Omega}_{l,n}$ and no more than one ``1'' in each column such that each antenna can only be selected once. Note that the sampling matrix will take $p_n$ values out of the $N_R$ samples. Thus, the sparsity rate is defined as 
	\begin{eqnarray}
		\iota=\frac{1}{N N_R}\sum_{n=1}^N p_n,   
	\end{eqnarray}
	which represents the volume ratio between the partial data and the complete data. Note that $\mathbf{p}_{c,l}(n)\sim \mathcal{CN}\left(\mathbf{0},\mathbf{\Omega}_{l,n}\mathbf{R}_l\mathbf{\Omega}_{l,n}^\TT\right)$ because $\mathbf{y}_{c,l}(n)\sim \mathcal{CN}\left(\mathbf{0},\mathbf{R}_l\right)$.
	
	The maximum likelihood (ML) estimate of the covariance matrix based on the partial data can be formulated as 
	\begin{equation}
		\begin{split}\label{rhoML}
			\widehat{\mathbf{R}}_{\text{ML},l}=\mathop{\arg \max}_{\mathbf{R}_{l}}  \mathcal{L}_{\mathbf{p},l} \left(\mathbf{R}_{l}  \Big|\left\{\mathbf{p}_{c,l}(n)\right\}_{n=1}^N,\left\{\mathbf{\Omega}_{l,n}\right\}_{n=1}^N \right),
		\end{split}
	\end{equation}
	where 
	\begin{equation}
		\begin{split}
			&\mathcal{L}_{\mathbf{p},l} \left(\mathbf{R}_{l} \Big|\left\{\mathbf{p}_{c,l}(n)\right\}_{n=1}^N,\left\{\mathbf{\Omega}_{l,n}\right\}_{n=1}^N \right)\\
			&\propto -\sum_{n=1}^{N}p_n\log \pi - \sum_{n=1}^{N} \left[\log \det \left( \mathbf{\Omega}_{l,n}\mathbf{R}_{l}\mathbf{\Omega}_{l,n}^\TT  \right) +\mathbf{p}_{c,l}^\HH(n)\left( \mathbf{\Omega}_{l,n}\mathbf{R}_{l}\mathbf{\Omega}_{l,n}^\TT  \right)^{-1}\mathbf{p}_{c,l}(n)\right],
		\end{split}
	\end{equation}
	denotes the log-likelihood function  of $\widehat{\mathbf{R}}_{l}$ based on $\left\{\mathbf{p}_{c,l}(n)\right\}_{n=1}^N$. 
	However, it is difficult to solve (\ref{rhoML}) directly since a closed-form solution is not available. Moreover, a solution based on an exhausted grid searching in the unknown parameter space could be computationally prohibitive. This motivates us to consider an approximate estimation of $\mathbf{R}_{l}$.
	
	\subsection{Expectation-maximization Algorithm}
	The log-likelihood function of $\widehat{\mathbf{R}}_{l}$ based on the complete data is given as
	\begin{equation}
		\begin{split}
			\mathcal{L}_{\mathbf{y},l} \left(\widehat{\mathbf{R}}_{l} \right)=-N\left(N_R \log\pi + \log \det\left(\widehat{\mathbf{R}}_l\right)    +   \tr \left(\widehat{\mathbf{R}}_l^{-1} \widehat{\mathbf{R}}_{\text{SCM},l}\right)\right),
		\end{split}
	\end{equation}
	where 
	\begin{eqnarray}
		\widehat{\mathbf{R}}_{\text{SCM},l}
		\triangleq\frac{1}{N} \sum_{n=1}^{N} \mathbf{y}_{c,l}(n)\mathbf{y}_{c,l}^\HH(n). 
	\end{eqnarray}
	However, $\widehat{\mathbf{R}}_{\text{SCM},l}$ is not available because the complete data $\{\mathbf{y}_{c,l}(n)\}_{n=1}^{N}$ is unknown. To solve this issue, we adopt the EM algorithm, which was proposed to find an approximate ML estimation with incomplete data \cite{bishop2006pattern}. The EM algorithm has two steps, i.e., the expectation step (E-step) and the maximization step (M-step).
	
	\subsubsection{E-step}
	At the E-step of the $t$th iteration, instead of finding $\mathcal{L}_{\mathbf{y},l} \left(\widehat{\mathbf{R}}_{l} \right)$, we find its conditional expectation 
	\begin{eqnarray}\label{Qrhocond}
		\bar{\mathcal{L}}\left(\widehat{\mathbf{R}}_{l};\mathbf{\Phi}_l^{(t)}\right) &=& \mbox{E}\left[ -N\left(N_R \log\pi + \log \det\left(\widehat{\mathbf{R}}_l\right)    +   \tr \left(\widehat{\mathbf{R}}_l^{-1} \widehat{\mathbf{R}}_{\text{SCM},l} \right)\right) \big|\mathbf{p}_{c,l}(n),\mathbf{\Omega}_{l,n},\widehat{\mathbf{R}}_{l}^{(t)} \right] \nonumber \\
		& =& -N\left(N_R \log\pi + \log \det\left(\widehat{\mathbf{R}}_l\right)    +   \tr \left(\widehat{\mathbf{R}}_l^{-1}\mathbf{\Phi}_l^{(t)}\right)\right)
	\end{eqnarray}
	where  
	$\mathbf{\Phi}_l^{(t)} =\frac{1}{N}\sum_{n=1}^{N} \mathbf{S}_n^{(t)}$ with
	\begin{equation}\label{Philt}
		\begin{split}
			\mathbf{S}_n^{(t)} \triangleq \mathbb{E} \left( \mathbf{y}_{c,l}(n)\mathbf{y}_{c,l}^\HH(n) \big|\mathbf{p}_{c,l}(n),\mathbf{\Omega}_{l,n},\widehat{\mathbf{R}}_{l}^{(t)} \right).
		\end{split}
	\end{equation}
	The following proposition gives the evaluation of $\mathbf{S}_n^{(t)}$.
	\begin{proposition}\label{lemmaCondExp}
		The conditional expectation for the covariance matrix of $\mathbf{y}_{c,l}(n)$ is given as
		\begin{equation} \label{conexp}
			\begin{split}
				\mathbf{S}_n^{(t)}=\left(\mathbf{\Omega}_{l,n}^\TT \mathbf{p}_{c,l}(n) + \overline{\mathbf{\Omega}}_{l,n}^\TT \mathbf{k}_{c,l}^{(t)}(n)  \right)\left(\mathbf{\Omega}_{l,n}^\TT \mathbf{p}_{c,l}(n) + \overline{\mathbf{\Omega}}_{l,n}^\TT \mathbf{k}_{c,l}^{(t)}(n)  \right)^\HH + \overline{\mathbf{\Omega}}_{l,n}^\TT \mathbf{\Psi}_{l,n}^{(t)} \overline{\mathbf{\Omega}}_{l,n},
			\end{split}
		\end{equation}
		where
		\begin{equation}\label{kclnt}
			\begin{split}
				\mathbf{k}_{c,l}^{(t)}(n)\triangleq\overline{\mathbf{\Omega}}_{l,n} \widehat{\mathbf{R}}_{l}^{(t)} \mathbf{\Omega}_{l,n}^\TT\left(\mathbf{\Omega}_{l,n} \widehat{\mathbf{R}}_{l}^{(t)} \mathbf{\Omega}_{l,n}^\TT\right)^{-1}\mathbf{p}_{c,l}(n),
			\end{split}
		\end{equation}
		\begin{equation}\label{psilnt}
			\begin{split}
				\mathbf{\Psi}_{l,n}^{(t)}\triangleq\overline{\mathbf{\Omega}}_{l,n} \widehat{\mathbf{R}}_{l}^{(t)}\overline{\mathbf{\Omega}}_{l,n}^\TT
				-\overline{\mathbf{\Omega}}_{l,n} \widehat{\mathbf{R}}_{l}^{(t)} \mathbf{\Omega}_{l,n}^\TT\left(\mathbf{\Omega}_{l,n} \widehat{\mathbf{R}}_{l}^{(t)} \mathbf{\Omega}_{l,n}^\TT\right)^{-1}\mathbf{\Omega}_{l,n} \widehat{\mathbf{R}}_{l}^{(t)} \overline{\mathbf{\Omega}}_{l,n}^\TT,
			\end{split}
		\end{equation}
		and $\overline{\mathbf{\Omega}}_{l,n}$ denotes the complement selection of $\mathbf{\Omega}_{l,n}$.
	\end{proposition}
	\emph{Proof}: See Appendix \ref{lemmaCondExpproof}.
	
	By substituting (\ref{conexp}) into (\ref{Qrhocond}), we can obtain $\bar{\mathcal{L}}\left(\widehat{\mathbf{R}}_{l};\mathbf{\Phi}_l^{(t)}\right)$. 
	
	\subsubsection{M-step}
	The M-step finds the update of $\widehat{\mathbf{R}}_{l}$ that maximizes $\bar{\mathcal{L}}\left(\widehat{\mathbf{R}}_{l};\mathbf{\Phi}_l^{(t)}\right)$.
	The update of $\widehat{\mathbf{R}}_{l}$ can be obtained by setting $\partial \bar{\mathcal{L}}\left(\widehat{\mathbf{R}}_{l};\mathbf{\Phi}_l^{(t)}\right)/\partial\widehat{\mathbf{R}}_{l}=\widehat{\mathbf{R}}_{l}^{-1}-\widehat{\mathbf{R}}_{l}^{-1}\mathbf{\Phi}_{l}^{(t)}\widehat{\mathbf{R}}_{l}^{-1}=
	\mathbf{0}, $
	which gives
	\begin{equation}\label{EMconv}
		\begin{split}
			\widehat{\mathbf{R}}_{l}^{(t+1)}=\mathbf{\Phi}_{l}^{(t)}.
		\end{split}
	\end{equation}
	However, $\mathbf{\Phi}_{l}^{(t)}$ can be ill-conditioned if the number of samples $N$ is smaller than the dimension $N_R$. 
	As a result, the inversion operation in (\ref{AMFresult}) may cause serious errors. In the following, we propose an unfolding method to solve this problem. 
	
	\subsection{EM-Net: Unfolded EM Algorithm}

	\subsubsection{Penalized EM Estimator}
	
	To tackle the above mentioned ill-conditioned issue, we consider adding a penalty term in $\bar{\mathcal{L}}\left(\widehat{\mathbf{R}}_{l};\mathbf{\Phi}_l^{(t)}\right)$ to improve the condition number of $\widehat{\mathbf{R}}_{l}$. It was shown in \cite{9686699} that the Kullback-Leibler (KL) divergence for Gaussian distributions, i.e., $	\mathcal{D}_{\text{KL}}\left(\widehat{\mathbf{R}}_{l}^{-1},\mathbf{I}\right)=\tr \left(\widehat{\mathbf{R}}_{l}^{-1} \right)-\log \det\left(\widehat{\mathbf{R}}_{l}^{-1}\right)-N_R,$
	can effectively constrain the condition number of $\widehat{\mathbf{R}}_{l}$. Thus, we adopt the KL divergence penalty and the penalized objective function can be given by 
	\begin{equation}\label{Qpena}
		\begin{split} 
			\bar{\mathcal{L}}_{pen}\left(\widehat{\mathbf{R}}_{l};\mathbf{\Phi}_l^{(t)}\right)=\bar{\mathcal{L}}\left(\widehat{\mathbf{R}}_{l};\mathbf{\Phi}_l^{(t)}\right)-\alpha_{l}^{(t)} N 
			\mathcal{D}_{\text{KL}}\left(\widehat{\mathbf{R}}_{l}^{-1},\mathbf{I}\right),
		\end{split}
	\end{equation}
	where $\alpha_{l}^{(t)}$ denotes the penalty coefficient. The maximizer of (\ref{Qpena}) gives the update of $\widehat{\mathbf{R}}_{l}$, i.e.,
	\begin{equation}\label{RlestShrink}
		\begin{split}
			\widehat{\mathbf{R}}_{l}^{(t+1)}=(1-\rho_{l}^{(t)})\mathbf{\Phi}_{l}^{(t)} +\rho_{l}^{(t)}\mathbf{I},
		\end{split}
	\end{equation}
	where 
	$\rho_{l}^{(t)}=\frac{\alpha_{l}^{(t)}}{1+\alpha_{l}^{(t)}}$. 
	(\ref{RlestShrink}) is a form of shrinkage estimation of the covariance matrix \cite{5484583,9686699}, in which $\rho_{l}^{(t)}$ is referred to as the shrinkage coefficient. The choice of $\rho_{l}^{(t)}$ has been discussed based on the prior distribution of the original signal in relevant works \cite{5484583,9686699}. However, due to the use of partial data in this paper, the closed-form solution of $\rho_{l}^{(t)}$ is difficult to obtain. In addition, $\rho_{l}^{(t)}$ changes in each iteration and the number of iterations is unpredictable, which makes the computational complexity of exhausted searching very high. Thus, we consider determining this parameter by the following unfolding method. 
	
	\subsubsection{EM-Net}
\begin{figure}[!t]
		\centering
		\includegraphics[width=3.21in]{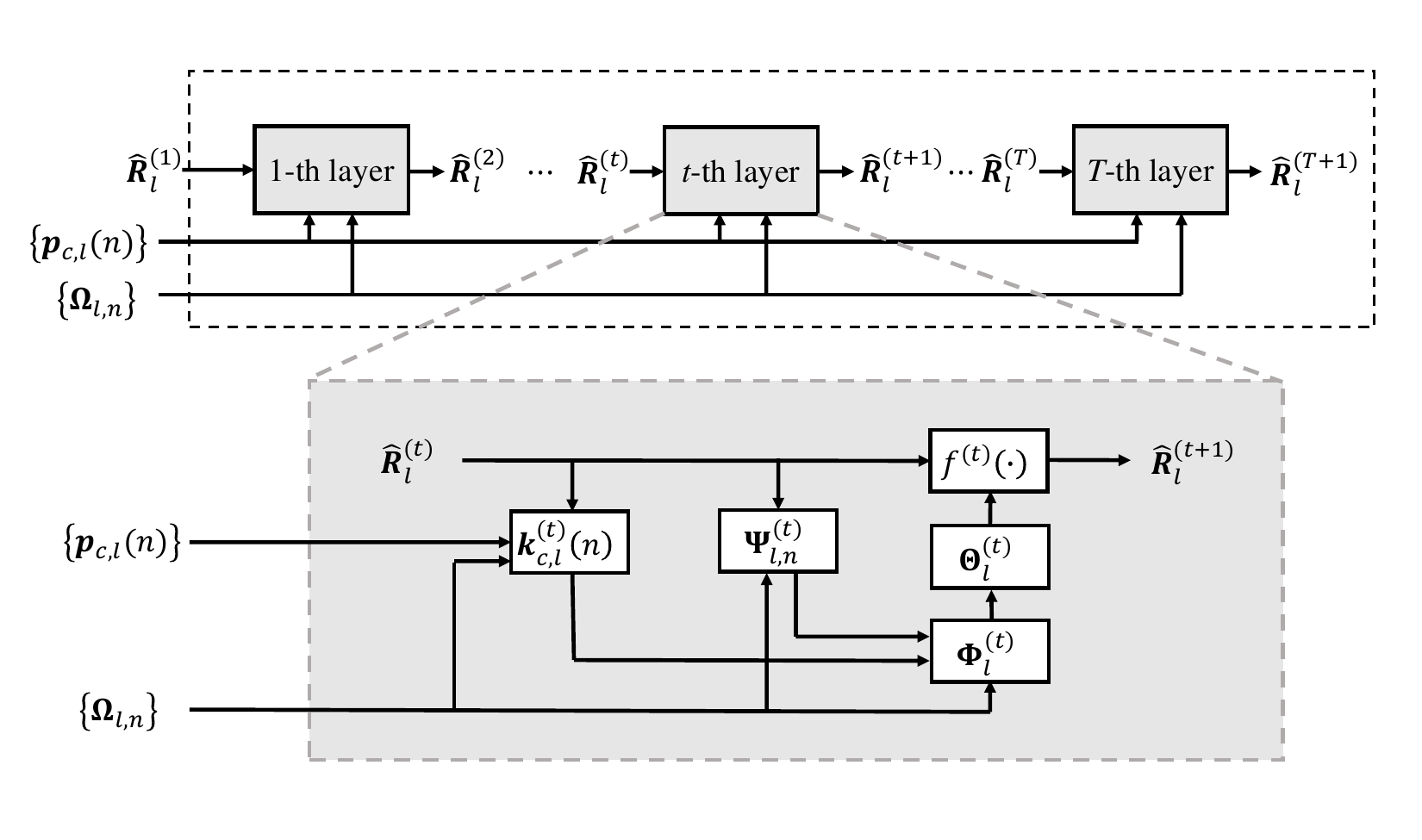}
		\caption{Diagram of the proposed EM-Net.}
		\label{fig_Network}
	\end{figure}

	The structure of the EM-Net is illustrated in Fig. \ref{fig_Network}, which is obtained by unfolding the EM detector and adding several trainable parameters.
	The network consists of $T$ cascaded layers sharing the same architecture but different trainable parameters. The input of the $t$th layer in the EM-Net is the output from the previous layer. The update in the $t$-th layer is shown in the lower part of Fig. \ref{fig_Network} with the detailed operations as follows
	\begin{equation}
			\mathbf{\Phi}_l^{(t)}=\frac{1}{N}\sum_{n=1}^{N} \mathbf{S}_n^{(t)},\;
		\mathbf{\Theta}_l^{(t+1)}=(1-\rho_{l}^{(t)})\mathbf{\Phi}_{l}^{(t)} +\rho_{l}^{(t)}\mathbf{I},
		\;\widehat{\mathbf{R}}_{l}^{(t+1)}=f_l^{(t)}\left(\mathbf{\Theta}_{l}^{(t+1)} ;\xi_{l}^{(t)} \right),
	\end{equation}
	where the divergence-free estimator $f_l^{(t)}\left(\cdot \right)$ is constructed by 
	\begin{equation}\label{DivFreeEst}
		\begin{split}
			f_l^{(t)}\left(\mathbf{\Theta}_{l}^{(t+1)} ;\xi_{l}^{(t)} \right)=(1-\xi_{l}^{(t)})\mathbf{\Theta}_{l}^{(t+1)} +\xi_{l}^{(t)}\widehat{\mathbf{R}}_{l}^{(t)}.
		\end{split}
	\end{equation}
	With the divergence-free estimator in (\ref{DivFreeEst}), the estimation results will not change dramatically after the $t$th layer \cite{7817805}.
	
	\begin{remark}
		The key difference between the EM and EM-Net algorithms is the learnable variables $\mathcal{S}_l=\{\rho_{l}^{(t)}, \xi_{l}^{(t)}\}$ in each layer. The learnable parameter $\rho_{l}^{(t)}$ controls the balance between estimation performance and the condition number. Another learnable parameter $\xi_{l}^{(t)}$ in the linear estimator $f_l^{(t)}\left(\cdot \right)$ plays an important role in constructing an appropriate divergence-free estimator. The original EM estimator can be interpreted as a special case of EM-net by setting $\rho_{l}^{(t)}=0$ and $\xi_{l}^{(t)}=0$.	By optimizing the learnable parameters in the training process, the estimation performance can be improved. 
	\end{remark}
	
	\begin{remark}
		Here, we evaluate the computational complexity of the proposed EM and EM-Net algorithms for one TMT. The complexity is dominated by the order of the number of complex-valued multiply operations. Recalling (\ref{Philt}), the complexity of SCM is $\mathcal{O}(N N_R^2)$. However, to achieve a satisfying  performance, a large number of samples are required, which leads to high hardware and power consumptions. Moreover, SCM is not suitable for the sparsely sampled data. The complexity of EM-Net is similar to that of EM because EM-Net has a similar structure to EM, but with some learnable parameters. Recalling (\ref{kclnt}) and (\ref{psilnt}), the complexity of obtaining $\mathbf{k}_{c,l}^{(t)}(n)$ and $\mathbf{\Psi}_{l,n}^{(t)}$ are dominated by the inverse operation of a $p_n \times p_n$ matrix, whose computational complexity is about $\mathcal{O}(p_n^3)$. The computational complexity for obtaining $\mathbf{\Phi}_l^{(t)}$ is about $\mathcal{O}(N N_R^2 +\sum_{n=1}^N p_n^3)$. Given EM-Net has $T$ layers, the overall computational computational complexity is $\mathcal{O}(T N N_R^2 +\sum_{n=1}^N T p_n^3)$. In general, the computational complexity of EM-Net is higher than that of SCM with the same number of samples. However, as shown in the simulation part, the EM-Net algorithm can significantly reduce the requirement of samples. It indicates that EM-Net can reduce the hardware and power consumption and system latency, which is attractive in real applications.
	\end{remark}
	
	\section{Simulation}
	In this section, we show the performance of the proposed networked sensing. Consider a PMN where the BSs are equipped with $N_T = 32$ antennas. The carrier frequency is set to 28 GHz and  $\beta=2$ \cite{akdeniz2014millimeter}. The noise power $\sigma^2$ is set as $-90$ dBm. We set 
	$\frac{1}{L}\sum_{i=1}^P\sum_{l=1}^{L}\sigma_{l,i}^2/\sigma^2=30$dB unless otherwise specified. 
	The channel between the BS and the $k$th UE is modeled as shown in (\ref{channelmodel}),
	where $\epsilon_{l,i}(n)\sim \mathcal{CN}(0,10^{-0.1(\kappa+\mu)})$ denotes the complex gain of the $i$-th path in the $n$-th subframe. Here, $\kappa$ is the path loss defined as $\kappa=a+10b\log_{10}(d)+\epsilon$ with $d$ denoting the distance between the BS and the $k$th UE and $\epsilon \sim \mathcal{CN}(0,\sigma_{\epsilon}^2)$ \cite{6834753}. Following \cite{6834753}, we set  $a=61.4$, $b=2$, and  $\sigma_{\epsilon}=5.8$dB. The channel is assumed to follow Rician fading, where the Rician factor is set as 7dB for the LoS component and 0dB for the NLoS component. For a given false alarm probability $P_{fa}$,  the detection threshold $\gamma_L$ is determined by (\ref{threshold}). Then, 100000 Monte-Carlo trials are performed to obtain $P_d$.

	\begin{figure}[!t]
		\centering
		\includegraphics[width=3.01in]{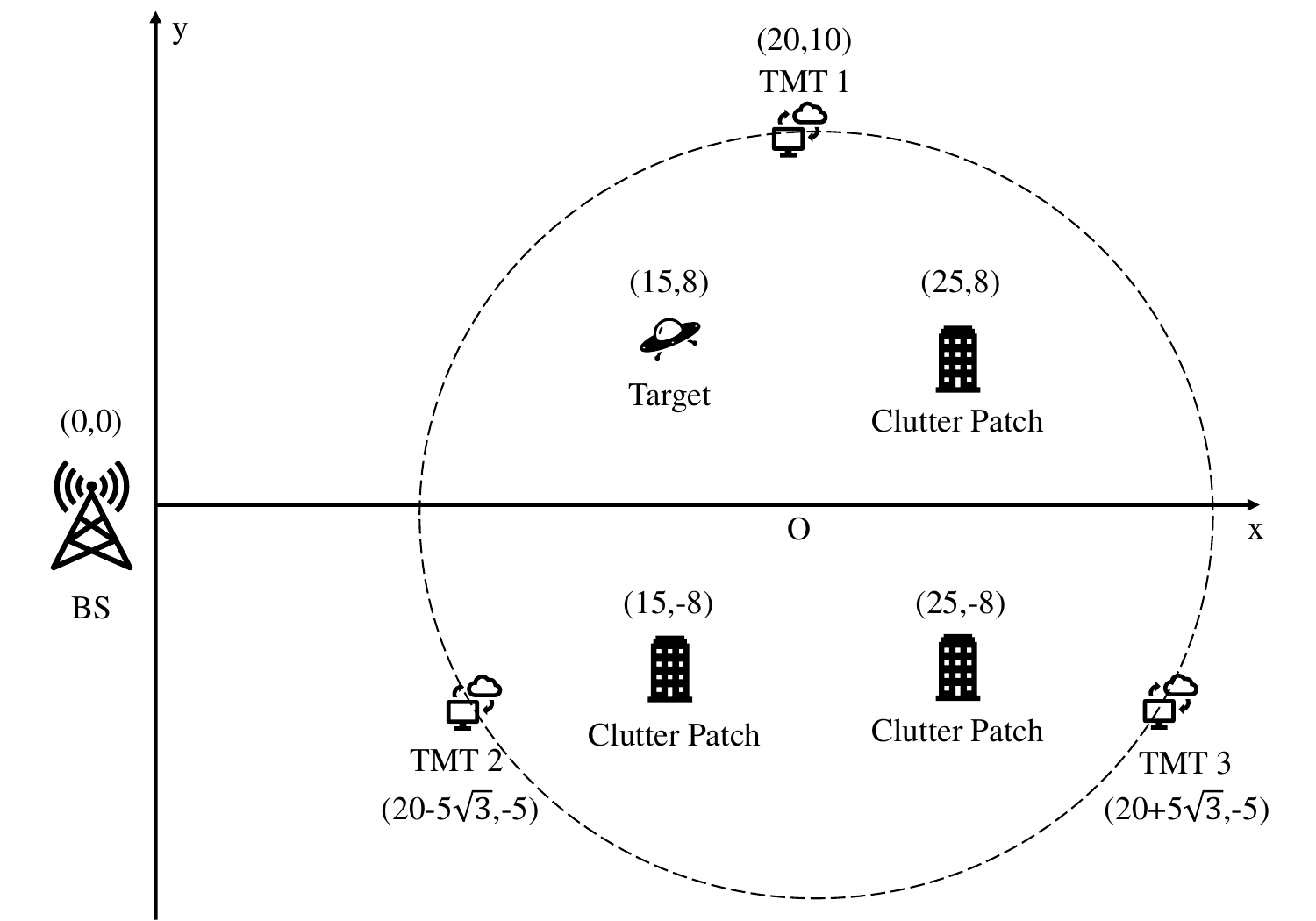}
		\caption{The illustration of the simulation scenario.}
		\label{fig_sysill}
	\end{figure}
	
	\begin{figure}[!t]
		\centering
		\includegraphics[width=3.01in]{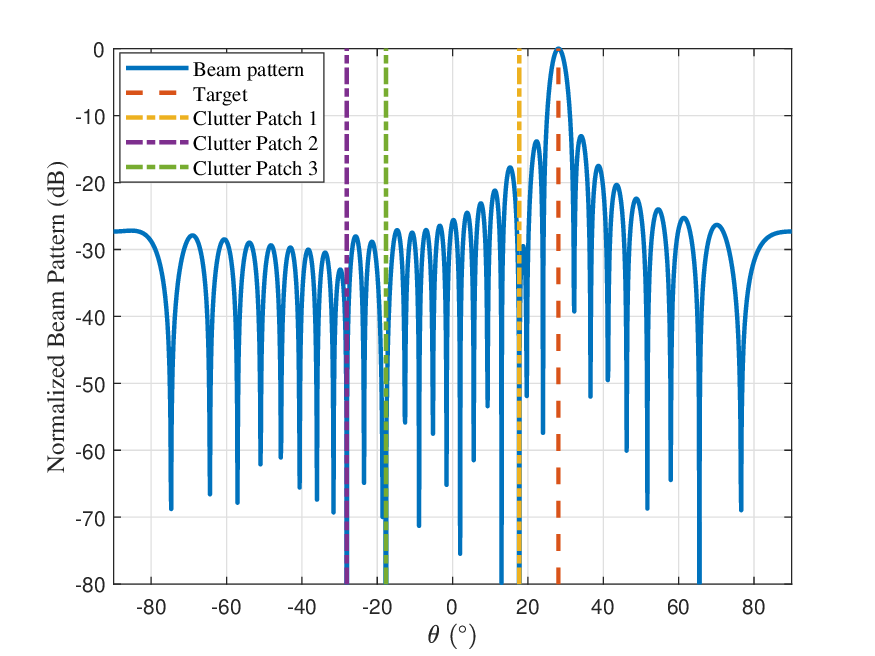}
		\caption{The normalized transmitting beam pattern.}
		\label{fig_bp}
	\end{figure}
	\subsection{Detection Performance}
	
	Consider a PMN with one BS and 3 TMTs, as illustrated in Fig. \ref{fig_sysill}.
	Assume that there are one target and 3 clutter patches in the environment. 
	The coordinates of the BS, the target, the clutter patches and the TMTs are respectively given as $(0,0)$, $(15,8)$, $\{(25,8),(15,-8),(25,-8)\}$, and $\{(20,10),(20-5\sqrt{3},-5),(20+5\sqrt{3},-5)\}$ where all coordinates are in meters.
	The AOD and AOA of the ST, UEs and clutter patches are determined based on the geometric locations. For each abscissa, 5000 Monte-Carlo trials are performed. 
	
	We first show the beam pattern given by $P(\theta)=|\mathbf{f}_{\perp}\mathbf{a}_T(\theta)|^2, \theta \in [-\frac{\pi}{2},\frac{\pi}{2}]$, which measures the transmitted power at the direction $\theta$. For clarity, the beam pattern is normalized by 
	$P_N(\theta)=\frac{P(\theta)}{\max_{\theta \in \Theta}P(\theta)}.$
	From Fig. \ref{fig_bp}, we can observe that the power transmitted to the clutter patches is lower than $-40$ dB, indicating that the sensing signal will not significantly affect the clutter patches. 
	
	\begin{figure}[!t]
		\centering
		\includegraphics[width=3.01in]{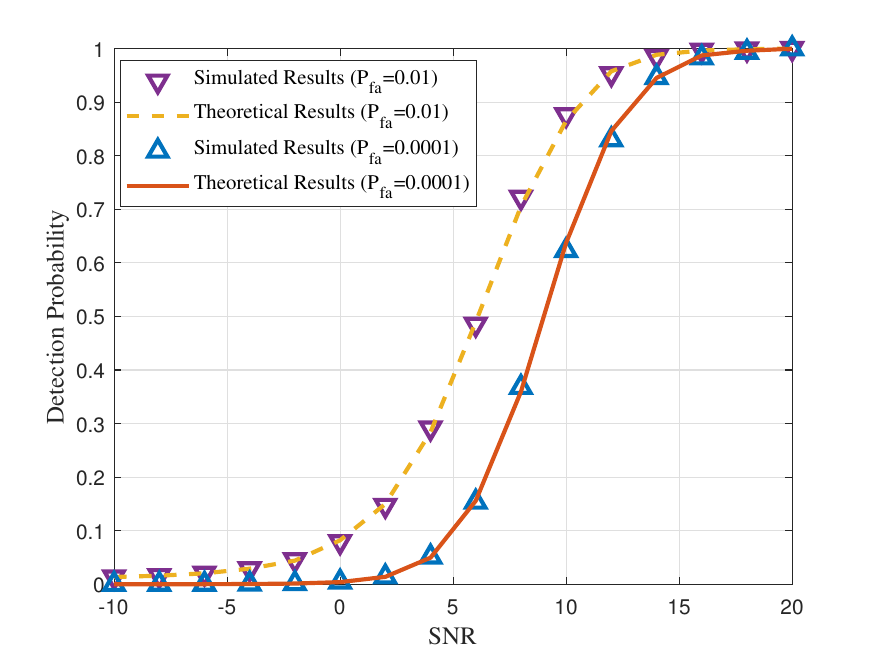}\label{fig_Pd_accuracy_test}\
		\caption{Accuracy for the Detection Probability $P_d^{(L)}$.}
		\label{fig_fitting}
	\end{figure}
	
	
	Fig. \ref{fig_fitting} shows the accuracy of the theoretical results regarding the detection threshold and the detection probability, respectively.  The legend ``Simulated Results'' indicates the results obtained by Monte-Carlo trials, while ``Theoretical Results'' represents the detection threshold and the detection probability obtained by (\ref{threshold}) and (\ref{AMFl1}), respectively. It can be observed that the theoretical results match the simulation results very well.

	\begin{figure}[!t]
		\centering
		\subfloat[]{\includegraphics[width=3.01in]{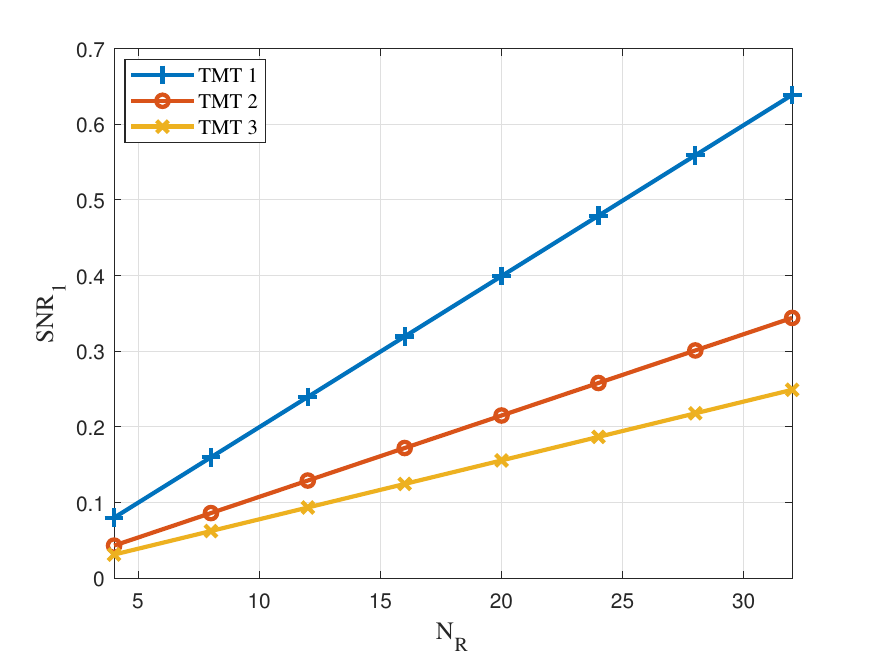}}\
		\subfloat[]{\includegraphics[width=3.01in]{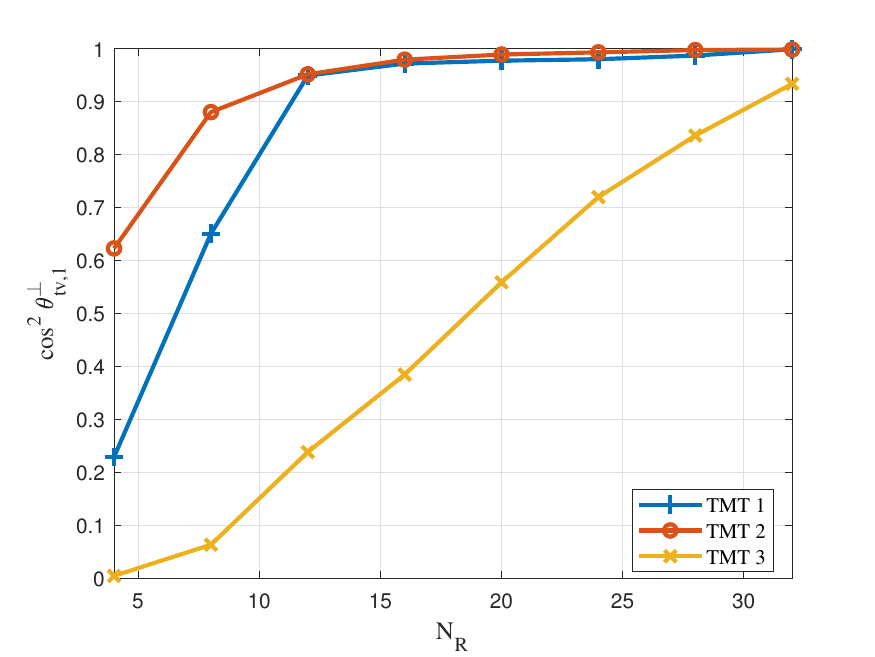}}\
		\caption{Effect of the number of antennas $N_R$. (a) $\text{SNR}_l$ (b) $\cos^2 \theta_{tv,l}^\perp$.}
		\label{fig_N_R_0}
	\end{figure}
	\subsection{Array Gain and Angular Resolution}
	It follows from (\ref{vsproj}) that the number of antennas $N_R$ influences  $\mu_l^2$ through SNR$_l$ and $\cos^2 \theta_{tv,l}^\perp$.   
	Parts (a) and (b) of Fig. \ref{fig_N_R_0} show the impact of $N_R$ on SNR$_l$ and $\cos^2 \theta_{tv,l}^\perp$, respectively, where we set  $C_g=10$ and $P_T=1$, and the other settings are the same as Fig. \ref{fig_fitting}. 
	It can be observed that SNR$_l$ grows linearly with $N_R$ due to the array gain. However, the improvement of $\cos^2 \theta_{tv,l}^\perp$ depends on the relative locations of the target and the clutter patches as shown in Fig. \ref{fig_sysill}. For example, $\cos^2 \theta_{tv,2}^\perp$  for TMT-2 increases quickly due to the clear link. However, $\cos^2 \theta_{tv,3}^\perp$ for TMT-3 grows very slowly due to the two nearby clutter patches. 
	
	\subsection{Macro-Diversity}	
	Next, we show the effect of the number of TMTs. For that purpose, we assume there are in total $Q$ TMTs available and show the performance when $L$ of them are selected. We set  $C_g=10$ and $P_T=0.5$, and the locations of the clutter patches are the same as those in Fig. \ref{fig_sysill}. The $Q$ TMTs are evenly located on a circle $\mathcal{O}$ with a radius of $10$m, where the coordinates of the $i$th TMT are given as $\left[20+10\sin(2\pi\frac{i-1}{Q}),10\cos(2\pi\frac{i-1}{Q})\right], i=1,2,\cdots,Q$. The target is randomly generated within the circle $\mathcal{O}$. For each abscissa, 10000 Monte-Carlo trials are performed and we set $P_{fa}=0.01$.

\begin{figure}[!t]
		\centering
		\subfloat[]{\includegraphics[width=3.01in]{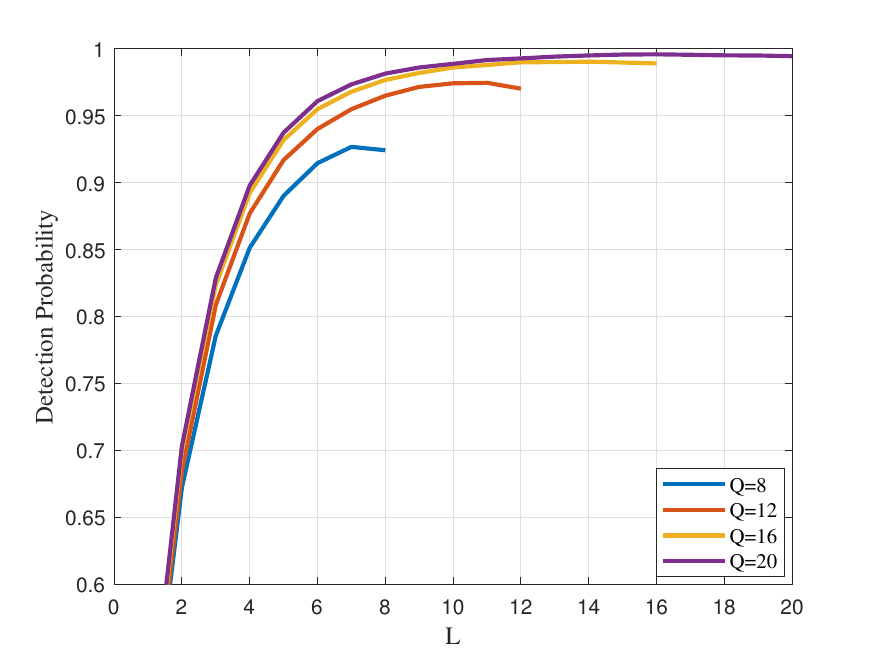}\label{fig_LfixedQ}}\
		\subfloat[]{\includegraphics[width=3.01in]{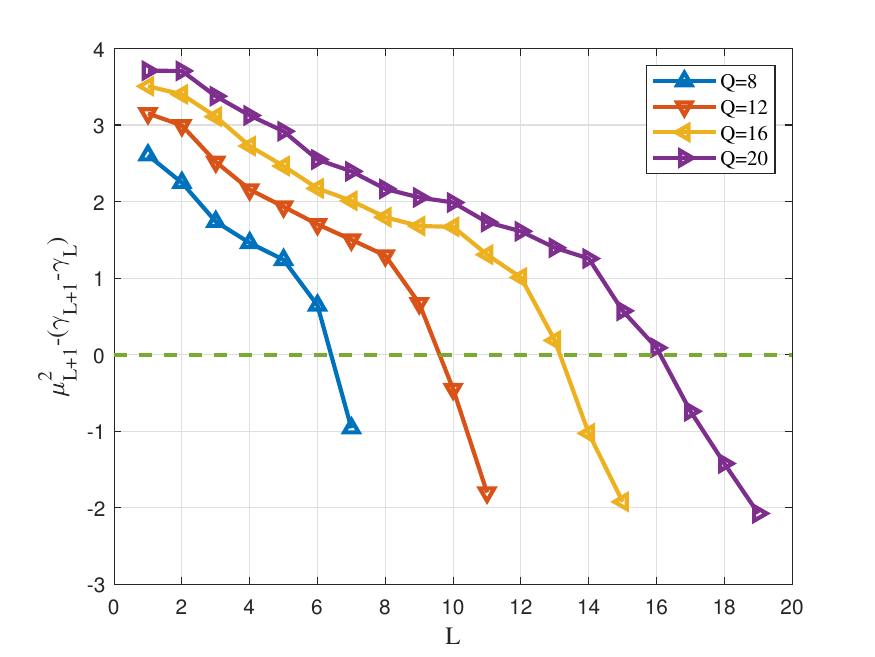}\label{fig_LfixedQ_cond}}\
		\caption{Detection Probability versus the number of selected TMTs $L$ and the number of  deployed TMTs $Q$. (a) The effect of $L$ with fixed $Q$; (b) The difference between the increment of $\zeta_L$ and $\gamma_L$:  $\mu_{L+1}^2-(\gamma_{L+1}-\gamma_{L})$. }
		\label{fig_zetavsLQ3}
	\end{figure}

	Fig. \ref{fig_zetavsLQ3} shows the detection probability $P_d$ when the best $L$ out of $Q$ TMTs are selected. In particular, we calculate $\{\mu_l^2\}_{l=1}^Q$ for all TMTs and arrange them in the descending order, i.e., $\mu_{i^{(1)}}^2 \geq \mu_{i^{(2)}}^2 \geq \cdots \mu_{i^{(Q)}}^2$. Then the $L$ TMTs with the highest $\mu_l^2$ are selected. We have two observations. First, for a given $L$, $P_d$ is an non-decreasing function of $Q$ due to the selection diversity. However, for a given $Q$, $P_d$ is not a monotonic increasing function of $L$. In fact, $P_d$ will first increase, then stabilize, and finally decrease.  This agrees with the discussion in Remark \ref{remark2}. 
	
	Next, we illustrate the performance of the proposed TMT selection algorithm. Assume $L$ TMTs have been selected.
	Fig. \ref{fig_LfixedQ_cond} shows the difference between the increment of $\zeta_L$ and $\gamma_L$, i.e., $\mu_{L+1}^2-(\gamma_{L+1}-\gamma_{L})$. According to \emph{Proposition \ref{mono00}}, adding the new TMT will benefit $P_d$, if $\mu_{L+1}^2\geq \gamma_{L+1}-\gamma_{L}$. We can observe from Fig. \ref{fig_LfixedQ_cond} that $\mu_{L+1}^2-(\gamma_{L+1}-\gamma_{L})$ is positive when $L$ is small, but as $L$ increases, $\mu_{L+1}^2-(\gamma_{L+1}-\gamma_{L})$ decreases and then becomes negative.
	For example, when $Q=12$, the cross-zero point is between $L=9$ and $L=10$, i.e., $\mu_{11}^2<(\gamma_{11}-\gamma_{10})$. Therefore, Algorithm \ref{alg_selection} will stop at $L=10$. But, the optimal $P_d$ is achieved at $L=11$ as shown in Fig. \ref{fig_LfixedQ}. This is because the conditions in \emph{Proposition \ref{mono00}} are sufficient but not necessary. As a result, the proposed algorithm gives a conservative but relatively accurate estimation for the optimal number of TMT. Such a conservative estimation is preferred from the system complexity point of view.
	
	\subsection{Covariance Matrix Estimation}
	Next, we show the performance of the proposed EM-Net algorithm for CE. Unless otherwise specified, the number of the antennas at one TMT and the number of communication subframes in the CE period are set as  $N_R=16$ and $N=50$, respectively. We set the number of TMTs and clutter patches to be $L=3$ and $P=3$. The coordinates of the TMTs and clutter patches are the same as those in Sec. V.A, and the target is randomly generated in the circle $\mathcal{O}$. Here, we set $P_{fa}=0.01$.
	
	For the training process, the learnable parameters are optimized by the stochastic gradient descent method. In our experiments, the loss function used for training is selected as $	f_{loss}=\left(\frac{1}{N_{\text{layer}}}\sum_{i=1}^{N_{\text{layer}}} \text{SL} \left(\mathbf{R}^{(i)},\widehat{\mathbf{R}}^{(i)},\mathbf{a}_t\right)\right)^{-1}$,
	where $N_{\text{layer}}$ denotes the number of layers in the training process, and $\text{SL} \left(\mathbf{R},\widehat{\mathbf{R}},\mathbf{a}_t\right)=\frac{\left(\mathbf{a}_t^\HH  \widehat{\mathbf{R}}^{-1} \mathbf{a}_t\right)^2}{\left(\mathbf{a}_t^\HH  \mathbf{R}^{-1} \mathbf{a}_t\right)\left(\mathbf{a}_t^\HH  \widehat{\mathbf{R}}^{-1}\mathbf{R}\widehat{\mathbf{R}}^{-1} \mathbf{a}_t\right)}$
	denotes the SCNR loss of the covariance estimation, i.e., the ratio between the SCNR with the estimated covariance matrix and that with the real covariance matrix, which is widely used to measure the performance of covariance estimation in radar detection \cite{1263229,4101326,9052470,9686699}. 
	The smaller the SCNR loss is, the better the detection performance will be. The number of layers is set as $N_{\text{layer}}=10$. The number of batches for training process is set as $N_{\text{batch}}=1500$.
	The batch size for each iteration is set as the total number of the communication subframes. We compare the performance of the EM and EM-Net estimators with the classical SCM estimator which utilizes the complete data.
	
	\subsubsection{Convergence Performance}
	\begin{figure}[!t]
		\centering
		\includegraphics[width=3.01in]{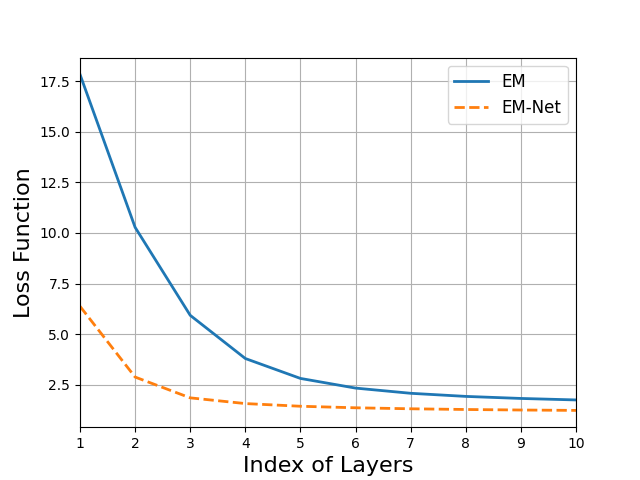}
		\caption{Loss function of the proposed EM and EM-Net estimators versus the number of layers.}
		\label{fig_Conv}
	\end{figure}
	First, we illustrate the convergence of the EM and EM-Net estimators. It can be observed from Fig. \ref{fig_Conv} that the EM and EM-Net detectors converge within 10 and 5 layers (iterations), respectively. Meanwhile, the EM-Net detector can achieve a lower training loss. 
	
	\subsubsection{Effect of Sample Size for Clutter Estimation}
	\begin{figure}[!t]
		\centering 
		\includegraphics[width=3.01in]{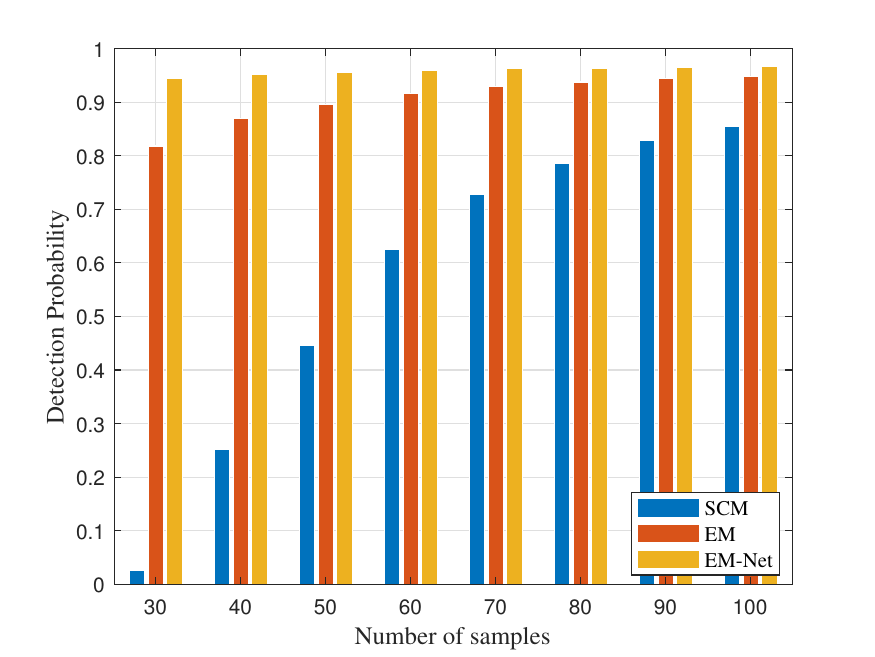}\label{fig_DPvsNumBatch}
		\caption{Detection probability versus the sample sizes for clutter estimation.}
		\label{fig_batch}
	\end{figure}
	
	Fig. \ref{fig_batch} shows the detection probability versus the sample size for clutter estimation. We set $N_R=16$, $\text{SNR}=10$ dB, and $\iota=0.5$ for all TMTs, while SCM requires the complete data. For each abscissa, 2000 Monte-Carlo trials are performed. It can be observed that the detection performance will improve as the sample size increases. Furthermore, EM-Net outperforms EM which achieves a better detection performance than SCM, and the performance gap is larger with less samples.

	\subsubsection{Effect of the Sparsity Rate}
	\begin{figure}[!t]
		\centering
		\includegraphics[width=3.01in]{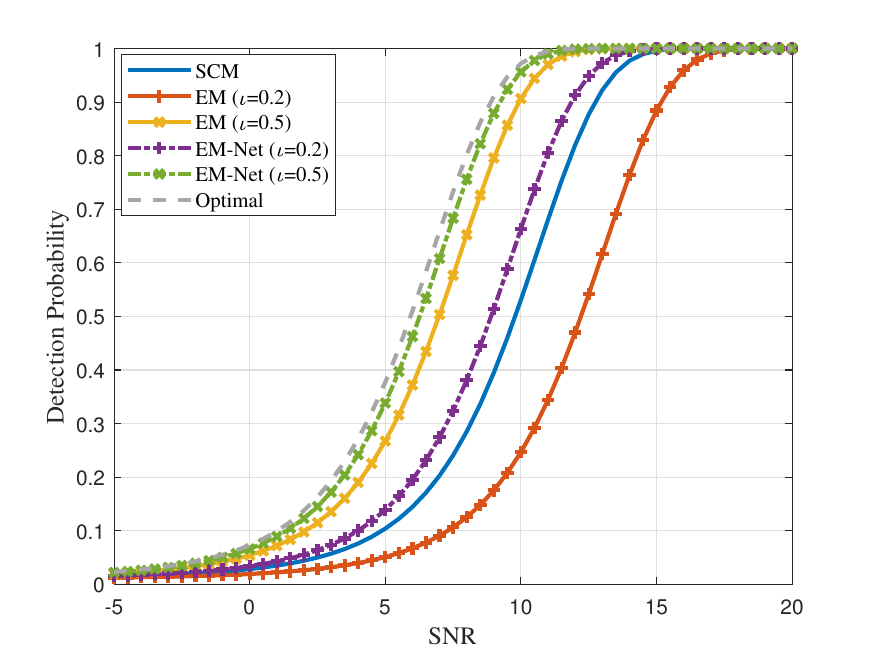}
		\caption{Detection Probability under different sparsity rate.}
		\label{fig_srvsPD}
	\end{figure}
	Fig. \ref{fig_srvsPD} depicts the detection probability versus SNR under different sparsity rates. The legend ``Optimal'' indicates the detection performance with the real covariance matrix.
	For each abscissa, 2000 Monte-Carlo trials are performed. From Fig. \ref{fig_srvsPD}, we can observe that EM-Net outperforms EM for different sparsity rates, indicating that the learnable parameters can improve the estimation performance.  Meanwhile, the detection performance of SCM is between EM and EM-Net with $\iota=0.2$, which shows that the proposed EM-Net can achieve better estimation performance with less data samples.
	
	\section{Conclusion}
	This paper investigated networked sensing in PMNs with the presence of clutter. A networked detector was developed to exploit the macro-diversity from multiple SNs, together with the array gain and higher angular resolution by multiple receive antennas, whose impact on sensing performance was investigated by theoretical analysis. It was shown that, although multiple SNs provide marco-diversity, the detection probability is not a monotonic increasing function of the number of activated SNs. A sufficient condition for one more SN's contribution to be positive was derived, with which a SN selection algorithm was proposed. Different from communication, although multiple receive antennas bring array gain and higher angular resolution, they no longer provide diversity gain for sensing because only the LoS component is used. To improve the efficiency of clutter estimation, an unfolded EM algorithm was proposed where only the partial data are required for estimating the clutter covariance. In summary, networked  sensing brings unprecedented opportunities to exploit the well-developed infrastructure of cellular networks for sensing purposes, but at the same time faces serious challenges in interference management with stringent computational and latency constraints, due to the collaboration of distributed nodes. This work revealed the advantage of networked sensing together with the impact of several key network parameters, and demonstrated the high efficiency of machine learning empowered clutter estimation algorithms.  
	\appendices

	\section{Proof of Proposition \ref{Prop1}}\label{Prop1proof}
	Before the proof, we define an auxiliary function $ \mathcal{F}_v\left(a,b\right)=\frac{\mathcal{Q}_{v+1}(\sqrt{a},\sqrt{b})}{\mathcal{Q}_v(\sqrt{a},\sqrt{b})}$ and introduce some important theorems as follows.
	\begin{theorem}[Monotonicity,{\cite[Theorem 1]{sun2010monotonicity}}]\label{mono00}
		The generalized Marcum Q-function $\mathcal{Q}_v(a,b)$ is strictly
		increasing in $v$ and $a$ when $a\geq 0$ and $b,v>0$, and is
		strictly decreasing in $b$ when $a,b\geq 0$ and $v>0$.
	\end{theorem}
	
	\begin{theorem}\label{Qderivative}
		The derivative of $\log {\mathcal{Q}_{v}(\sqrt{a},\sqrt{b})}$ with respect to $a$ and $b$ are, respectively, given by
			\begin{equation}\label{derivaab}
				\begin{split}
					\frac{\partial \log {\mathcal{Q}_{v}(\sqrt{a},\sqrt{b})}}{\partial a}=
					\frac{1}{2}\frac{\mathcal{Q}_{v+1}(\sqrt{a},\sqrt{b})}{\mathcal{Q}_{v}(\sqrt{a},\sqrt{b})}-\frac{1}{2},\;\frac{\partial \log {\mathcal{Q}_{v}(\sqrt{a},\sqrt{b})}}{\partial b}=
					\frac{1}{2}\frac{\mathcal{Q}_{v-1}(\sqrt{a},\sqrt{b})}{\mathcal{Q}_{v}(\sqrt{a},\sqrt{b})}-\frac{1}{2}.
				\end{split}
			\end{equation}
	\end{theorem}
	
	\emph{Proof:} (\ref{derivaab}) 
	can be directly derived by the chain rule as follows:
\begin{subequations}
	\begin{equation}
		\begin{split}
			&\frac{\partial \log {\mathcal{Q}_{v}(\sqrt{a},\sqrt{b})}}{\partial a}=\frac{\partial \log {\mathcal{Q}_{v}(\sqrt{a},\sqrt{b})}}{\partial \mathcal{Q}_{v}(\sqrt{a},\sqrt{b})}\cdot\frac{\partial\mathcal{Q}_{v}(\sqrt{a},\sqrt{b})}{\partial\sqrt{a}}\cdot\frac{\partial\sqrt{a}}{\partial a}\\
			&=\frac{1}{ \mathcal{Q}_{v}(\sqrt{a},\sqrt{b})} \left[\sqrt{a} \left(\mathcal{Q}_{v+1}(\sqrt{a},\sqrt{b})-\mathcal{Q}_{v}(\sqrt{a},\sqrt{b})\right)\right] \frac{1}{2\sqrt{a}}=\frac{1}{2}\frac{\mathcal{Q}_{v+1}(\sqrt{a},\sqrt{b})}{\mathcal{Q}_{v}(\sqrt{a},\sqrt{b})}-\frac{1}{2}.
		\end{split}
	\end{equation}
	\begin{equation}
		\begin{split}
			&\frac{\partial \log {\mathcal{Q}_{v}(\sqrt{a},\sqrt{b})}}{\partial b}=\frac{\partial \log {\mathcal{Q}_{v}(\sqrt{a},\sqrt{b})}}{\partial \mathcal{Q}_{v}(\sqrt{a},\sqrt{b})}\cdot\frac{\partial\mathcal{Q}_{v}(\sqrt{a},\sqrt{b})}{\partial\sqrt{b}}\cdot\frac{\partial\sqrt{b}}{\partial b}\\
			&=\frac{1}{ \mathcal{Q}_{v}(\sqrt{a},\sqrt{b})} \left[\sqrt{b} \left(\mathcal{Q}_{v-1}(\sqrt{a},\sqrt{b})-\mathcal{Q}_{v}(\sqrt{a},\sqrt{b})\right)\right] \frac{1}{2\sqrt{b}}=\frac{1}{2}\frac{\mathcal{Q}_{v-1}(\sqrt{a},\sqrt{b})}{\mathcal{Q}_{v}(\sqrt{a},\sqrt{b})}-\frac{1}{2}.
		\end{split}
	\end{equation}
\end{subequations}
	
	\begin{theorem}\label{auximono00}
		The auxiliary function $b \mapsto \mathcal{F}_v\left(a,b\right)$ is monotonically decreasing for $a \geq b > 0$.
	\end{theorem}
	
	\emph{Proof:}  When $a\geq 0$ and $b,v>0$,  $\mathcal{Q}_v(a,b)$ is strictly
	increasing in $v$ \cite[Theorem 1]{sun2010monotonicity}, and we have $\frac{\mathcal{Q}_{v}(\sqrt{a},\sqrt{b})}{\mathcal{Q}_{v+1}(\sqrt{a},\sqrt{b})}<1$.
	Meanwhile, in view of \cite[Theorem 3.1(b)]{sun2008inequalities}, the function $v \mapsto \mathcal{Q}_{v+1}(\sqrt{a},\sqrt{b})-\mathcal{Q}_{v}(\sqrt{a},\sqrt{b})$ is strictly decreasing on $(0,+\infty)$ for $a \geq b > 0$. Thus, we have $$	\mathcal{Q}_{v+1}(\sqrt{a},\sqrt{b})-\mathcal{Q}_{v}(\sqrt{a},\sqrt{b}) < \mathcal{Q}_{v}(\sqrt{a},\sqrt{b})-\mathcal{Q}_{v-1}(\sqrt{a},\sqrt{b}).$$
	We know that $\frac{x}{y}>\frac{x+u}{y+v}$ for all $0<y<x$ and $0<u<v$. Thus, invoking $x=\mathcal{Q}_{v}(\sqrt{a},\sqrt{b})$, $y=\mathcal{Q}_{v-1}(\sqrt{a},\sqrt{b})$, $u=\mathcal{Q}_{v+1}(\sqrt{a},\sqrt{b})-\mathcal{Q}_{v}(\sqrt{a},\sqrt{b})$, and $v=\mathcal{Q}_{v}(\sqrt{a},\sqrt{b})-\mathcal{Q}_{v-1}(\sqrt{a},\sqrt{b})$, we have
	\begin{equation}\label{QvQv1mono}
		\begin{split}
			\mathcal{F}_v\left(a,b\right)=\frac{\mathcal{Q}_{v}(\sqrt{a},\sqrt{b})+\left(\mathcal{Q}_{v+1}(\sqrt{a},\sqrt{b})-\mathcal{Q}_{v}(\sqrt{a},\sqrt{b})\right)}{\mathcal{Q}_{v-1}(\sqrt{a},\sqrt{b})+\left(\mathcal{Q}_{v}(\sqrt{a},\sqrt{b})-\mathcal{Q}_{v-1}(\sqrt{a},\sqrt{b})\right)}< 
			\mathcal{F}_{v-1 }\left(a,b\right).
		\end{split}
	\end{equation}
	
	\begin{theorem}\label{increment0}
		For all $a,b\geq 0$, $\Delta a, \Delta b >0$, and $v>0$, we have
		\begin{subequations}
			\begin{equation}\label{aaa1}
				\begin{split}
					\log \mathcal{Q}_{v}(\sqrt{a+\Delta a},\sqrt{b})-\log \mathcal{Q}_{v}(\sqrt{a},\sqrt{b})\geq 
					\left(\frac{1}{2}\frac{\mathcal{Q}_{v+1}(\sqrt{a+\Delta a},\sqrt{b})}{\mathcal{Q}_{v}(\sqrt{a+\Delta a},\sqrt{b})}-\frac{1}{2}\right)\Delta a,
				\end{split}
			\end{equation}
			\begin{equation}\label{aaa2}
				\begin{split}
					\log \mathcal{Q}_{v}(\sqrt{a},\sqrt{b+\Delta b})-\log \mathcal{Q}_{v}(\sqrt{a},\sqrt{b})\geq 
					\left(\frac{1}{2}\frac{\mathcal{Q}_{v-1}(\sqrt{a },\sqrt{b+\Delta b})}{\mathcal{Q}_{v}(\sqrt{a },\sqrt{b+\Delta b})}-\frac{1}{2}\right)\Delta b.
				\end{split}
			\end{equation}
		\end{subequations}
	\end{theorem}
	
	\emph{Proof:} 
	Since $a\mapsto \mathcal{Q}_v(\sqrt{a},\sqrt{b})$ and $b\mapsto \mathcal{Q}_v(\sqrt{a},\sqrt{b})$ are both log-concave when $a,b \leq 0$ and $v>0$ \cite{sun2008inequalities,sun2010monotonicity}, the function $a \mapsto \log \mathcal{Q}_v(\sqrt{a},\sqrt{b})$ and $b \mapsto \log \mathcal{Q}_v(\sqrt{a},\sqrt{b})$ are both concave.
	We begin with proving  (\ref{aaa1}).
	According to the mean value theorem \cite{sahoo1998mean}, there exists a point $\xi$ in $(a,a+\Delta a)$ such that 
	\begin{equation}\label{mvt1}
		\begin{split}
			\log \mathcal{Q}_v(\sqrt{a+\Delta a},\sqrt{b})-\log \mathcal{Q}_v(\sqrt{a},\sqrt{b}) & = 	\frac{\partial \log {\mathcal{Q}_{v}(\sqrt{a},\sqrt{b})}}{\partial a}\Big|_{a=\xi} \cdot \Delta a.
		\end{split}
	\end{equation}
	Given the property of the concave function, we have
	\begin{equation}\label{mvt2}
		\begin{split}
			\frac{\partial \log {\mathcal{Q}_{v}(\sqrt{a},\sqrt{b})}}{\partial a}\Big|_{a=\xi} \geq \frac{\partial \log {\mathcal{Q}_{v}(\sqrt{a},\sqrt{b})}}{\partial a}\Big|_{a=a+\Delta a} .
		\end{split}
	\end{equation}	
	In view of (\ref{mvt1}) and (\ref{mvt2}), the inequality in (\ref{aaa1}) is proved based on \emph{Theorem \ref{Qderivative}}. Similarly, the inequality in (\ref{aaa2}) can be obtained.
	\hfill $\blacksquare$
	
	Next, we will prove \emph{Proposition \ref{Prop1}}. Given $P_d^{(L+1)}>0$ and $P_d^{(L)}>0$, we have
	\begin{equation}
		\begin{split}
			\log \frac{P_d^{(L+1)}}{P_d^{(L)}}=\log \mathcal{Q}_{L+1}\left(\sqrt{\zeta_{L+1}},\sqrt{\gamma_{L+1}}\right)-\log \mathcal{Q}_{L}\left(\sqrt{\zeta_{L}},\sqrt{\gamma_{L}}\right)=\mathcal{A}_1+\mathcal{A}_2+\mathcal{A}_3,
		\end{split}
	\end{equation}
	where 
	\begin{equation}
		\begin{split}
			&\mathcal{A}_1=\log \mathcal{Q}_{L+1}\left(\sqrt{\zeta_{L+1}},\sqrt{\gamma_{L+1}}\right)-\log \mathcal{Q}_{L+1}\left(\sqrt{\zeta_{L+1}},\sqrt{\gamma_{L}}\right),\\
			&\mathcal{A}_2=\log \mathcal{Q}_{L+1}\left(\sqrt{\zeta_{L+1}},\sqrt{\gamma_{L}}\right)-\log \mathcal{Q}_{L}\left(\sqrt{\zeta_{L+1}},\sqrt{\gamma_{L}}\right),\\
			&\mathcal{A}_3=\log \mathcal{Q}_{L}\left(\sqrt{\zeta_{L+1}},\sqrt{\gamma_{L}}\right)-\log \mathcal{Q}_{L}\left(\sqrt{\zeta_{L}},\sqrt{\gamma_{L}}\right).
		\end{split}
	\end{equation}
	Recalling (\ref{GammaDet0}),  $\mathcal{A}_1$, $\mathcal{A}_2$, and $\mathcal{A}_3$ represent the increment of detection probability with respect to the detection threshold  $\gamma_{L}$, the DOF of the decision statistic $L$, and the non-central parameter $\zeta_{L}$, respectively. By observing \emph{Theorem \ref{mono00}}, we have that $\mathcal{A}_2\geq 0$, $\mathcal{A}_3 \geq 0$, and $\mathcal{A}_1<0$. The first two inequalities hold because increasing the DOF and the non-central parameter of $\Gamma_L$ will increase the detection probability. However, a larger $\gamma_{L}$ will decrease $P_d^{(L)}$, which leads to $\mathcal{A}_1<0$.
	
	By applying \emph{Theorem \ref{increment0}}, we have
	\begin{equation} \label{A1}
		\begin{split}
			\mathcal{A}_1\geq 
			-\frac{1}{2}\left(1-\frac{1}{\mathcal{F}_L(\zeta_{L+1},\gamma_{L+1})}\right)\left(\gamma_{L+1}-\gamma_{L}\right),
		\end{split}
	\end{equation}
	\begin{equation} \label{A3}
		\begin{split}
			\mathcal{A}_3\geq \left(\frac{1}{2}\frac{\mathcal{Q}_{L+1}(\sqrt{\zeta_{L+1}},\sqrt{\gamma_{L}})}{\mathcal{Q}_{L}(\sqrt{\zeta_{L+1}},\sqrt{\gamma_{L}})}-\frac{1}{2}\right) \mu_{L+1}^2=
			\frac{1}{2}\left(\mathcal{F}_L(\zeta_{L+1},\gamma_{L})-1\right) \mu_{L+1}^2.
		\end{split}
	\end{equation}
	Recalling the conditions 1) and 2) in \emph{Proposition \ref{Prop1}}, we have
	$\zeta_{L+1} > \gamma_{L+1}>0$ and $\zeta_{L+1} > \gamma_{L}>0$, which means that  \emph{Theorem \ref{auximono00}} holds true for both $\mathcal{F}_L(\zeta_{L+1},\gamma_{L+1})$ and $\mathcal{F}_L(\zeta_{L+1},\gamma_{L})$.
	Applying  \emph{Theorem \ref{auximono00}} yields $\mathcal{F}_L(\zeta_{L+1},\gamma_{L+1}) < \mathcal{F}_L(\zeta_{L+1},\gamma_{L}).$
	Thus, we have
	\begin{equation}\label{flfl1}
		\begin{split}
			&\mathcal{F}_L(\zeta_{L+1},\gamma_{L})\mathcal{F}_L(\zeta_{L+1},\gamma_{L+1})-2\mathcal{F}_L(\zeta_{L+1},\gamma_{L+1})+1\geq
			\left(\mathcal{F}_L(\zeta_{L+1},\gamma_{L+1})-1\right)^2\geq 0,
		\end{split}
	\end{equation}
	By rearranging (\ref{flfl1}), we have 
	\begin{equation}\label{fcoef}
		\begin{split}
			1-\frac{1}{\mathcal{F}_L(\zeta_{L+1},\gamma_{L+1})}\leq \mathcal{F}_L(\zeta_{L+1},\gamma_{L})-1.
		\end{split}
	\end{equation}
	From  \emph{Theorem \ref{mono00}}, we have $\mathcal{F}_v(a,b)>1$, such that 
	$	1-\frac{1}{\mathcal{F}_L(\zeta_{L+1},\gamma_{L+1})} >0$ and $\mathcal{F}_L(\zeta_{L+1},\gamma_{L})-1>0$.
	Based on (\ref{cond2}) and (\ref{fcoef}), we can rewrite (\ref{A3}) as  $		\mathcal{A}_3 \geq \frac{1}{2}\left( 1-\frac{1}{\mathcal{F}_L(\zeta_{L+1},\gamma_{L+1})} \right) \left( \gamma_{L+1}-\gamma_{L} \right) = -\mathcal{A}_1,$
	which gives $\mathcal{A}_1+\mathcal{A}_3\geq 0$. Given $\mathcal{A}_2\geq 0$, we have 
	$	\log \frac{P_d^{(L+1)}}{P_d^{(L)}}=\mathcal{A}_1+\mathcal{A}_2+\mathcal{A}_3\geq 0.$
	It follows that $P_d^{(L+1)}>P_d^{(L)}$, which completes the proof. \hfill $\blacksquare$
	
	\section{Proof of Proposition \ref{lemmaCondExp}}\label{lemmaCondExpproof}
	First, we define $\mathbf{Q}_{l,n}=\left[
	\mathbf{\Omega}_{l,n}^\TT,\overline{\mathbf{\Omega}}_{l,n}^\TT\right]^\TT$, and it can be validated that $\mathbf{Q}_{l,n}^\TT\mathbf{Q}_{l,n}=\mathbf{I}$. 
	Then, we have
	\begin{equation}
		\label{Sn}
		\mathbf{S}_n^{(t)}
		=\mathbb{E} \left( \mathbf{y}_{c,l}(n)\mathbf{y}_{c,l}^\HH(n) \big|\mathbf{p}_{c,l}(n),\mathbf{\Omega}_{l,n},\widehat{\mathbf{R}}_{l}^{(t)} \right)
		=\mathbf{Q}_{l,n}^\TT \mathbf{\Xi}_n
		\mathbf{Q}_{l,n},
	\end{equation}
	where
	\begin{eqnarray}\label{Eqyyq}
		&&\mathbf{\Xi}_n\triangleq\mathbb{E} \left(\mathbf{Q}_{l,n} \mathbf{y}_{c,l}(n)\mathbf{y}_{c,l}^\HH(n)\mathbf{Q}_{l,n}^\TT \big|\mathbf{p}_{c,l}(n),\mathbf{\Omega}_{l,n},\widehat{\mathbf{R}}_{l}^{(t)} \right) \\
		&&=\left[\begin{matrix}
			\mathbf{p}_{c,l}\mathbf{p}_{c,l}^\HH & \mathbf{p}_{c,l}\mathbb{E} \left(\mathbf{y}_{c,l}^\HH(n)\overline{\mathbf{\Omega}}_{l,n}^\TT \big|\mathbf{p}_{c,l}(n),\mathbf{\Omega}_{l,n},\widehat{\mathbf{R}}_{l}^{(t)} \right)\\
			\mathbb{E} \left(\overline{\mathbf{\Omega}}_{l,n} \mathbf{y}_{c,l}(n) \big|\mathbf{p}_{c,l}(n),\mathbf{\Omega}_{l,n},\widehat{\mathbf{R}}_{l}^{(t)} \right)\mathbf{p}_{c,l}^\HH & \mathbb{E} \left(\overline{\mathbf{\Omega}}_{l,n} \mathbf{y}_{c,l}(n)\mathbf{y}_{c,l}^\HH(n)\overline{\mathbf{\Omega}}_{l,n}^\TT \big|\mathbf{p}_{c,l}(n),\mathbf{\Omega}_{l,n},\widehat{\mathbf{R}}_{l}^{(t)} \right)\\
		\end{matrix}\right]. \nonumber
	\end{eqnarray}
	Given $\overline{\mathbf{\Omega}}_{l,n} \mathbf{y}_{c,l}(n)$ follows a Gaussian distribution, we have \cite{kay1993fundamentals} 
	\begin{equation}\label{meanqy}
		\begin{split}
			\mathbb{E} \left(\overline{\mathbf{\Omega}}_{l,n} \mathbf{y}_{c,l}(n) \big|\mathbf{p}_{c,l}(n),\mathbf{\Omega}_{l,n},\widehat{\mathbf{R}}_{l}^{(t)} \right)=	\mathbf{k}_{c,l}^{(t)}(n),
		\end{split}
	\end{equation}
	\begin{equation}\label{covqy}
		\begin{split}
			&\mathbb{E} \left(\overline{\mathbf{\Omega}}_{l,n} \mathbf{y}_{c,l}(n)\mathbf{y}_{c,l}^\HH(n)\overline{\mathbf{\Omega}}_{l,n}^\TT \big|\mathbf{p}_{c,l}(n),\mathbf{\Omega}_{l,n},\widehat{\mathbf{R}}_{l}^{(t)} \right)=\mathbf{\Psi}_{l,n}^{(t)}+\mathbf{k}_{c,l}^{(t)}(n)\mathbf{k}_{c,l}^{(t),\HH}(n).
		\end{split}
	\end{equation}
	By substituting (\ref{meanqy}) and (\ref{covqy}) into (\ref{Eqyyq}) and then (\ref{Sn}) yields (\ref{conexp}). \hfill$\blacksquare$


\end{document}